\documentclass{article}
\usepackage{arxiv}
\usepackage[utf8]{inputenc} 
\usepackage[T1]{fontenc}    
\usepackage{hyperref}       
\usepackage{url}            
\usepackage{booktabs}       
\usepackage{amsfonts}       
\usepackage{nicefrac}       
\usepackage{microtype}      
\usepackage{lipsum}
\usepackage[flushleft]{threeparttable}
\usepackage{xr-hyper}
\usepackage{hyperref}
\usepackage[detect-all]{siunitx}
\usepackage{import,gensymb}
\usepackage{colortbl}
\usepackage{booktabs}
\usepackage{array,multirow}
\usepackage{algorithm}
\usepackage{algorithmic}
\usepackage{graphicx}
\usepackage{subfigure}
\usepackage{epstopdf}
\usepackage{float}
\graphicspath{{Figures/}}
\title{Revisit Geophysical Imaging in A New View of Physics-informed Generative Adversarial Learning}

\author{
  Fangshu Yang \\
  School of Mathematics\\
  Harbin Institute of Technology\\
  \texttt{yfs2016@hit.edu.cn} \\
   \And
  Jianwei Ma \\
  School of Earth and Space Sciences\\
  Peking University\\
  \texttt{jwm@pku.edu.cn} \\
}

\begin{document}
\maketitle
\begin{abstract}
Seismic full waveform inversion~(FWI) is a powerful geophysical imaging technique that produces high-resolution subsurface models by iteratively minimizing the misfit between the simulated and observed seismograms. Unfortunately, conventional FWI with least-squares function suffers from many drawbacks such as the local-minima problem and computation of explicit gradient. It is particularly challenging with the contaminated measurements or poor starting models. Recent works relying on partial differential equations and neural networks show promising performance for two-dimensional FWI. Inspired by the competitive learning of generative adversarial networks, we proposed an unsupervised learning paradigm that integrates wave equation with a discriminate network to accurately estimate the physically consistent models in a distribution sense. Our framework needs no labelled training data nor pretraining of the network, is flexible to achieve multi-parameters inversion with minimal user interaction. The proposed method faithfully recovers the well-known synthetic models that outperforms the classical algorithms. Furthermore, our work paves the way to sidestep the local-minima issue via reducing the sensitivity to initial models and noise.
\end{abstract}


\section{Introduction}

Seismic velocity model is of great importance for the seismic exploration, especially the migration, imaging and interpretation processes. Accurate estimation of velocity model plays an essential role in the investigation of earth's structures and natural resource. In the last decades, such velocity information can be derived by some classical imaging methods such as velocity analysis~\cite{berkhout1997pushing}, tomography~\cite{iyer1993seismic}, and full waveform inversion~(FWI)~\cite{tarantola1984inversion,virieux2009overview,brossier2009seismic}. Compared to other traditional techniques, FWI can obtain high-resolution inversion results by iteratively matching the simulated seismograms, which obtained by numerical simulation based on the forward modeling equation, and the observed seismograms. Usually, the minimization of objective function of FWI is solved by adjoint-state method~\cite{plessix2006review}, in which the gradients of the cost with respect to~(w.r.t.) velocity model is computed explicitly. This leads to challenges and inconvenience when faced with the changes of the forward operator. Moreover, classical FWI is quite effective when a relatively good initial model and optimized gradient-based algorithm are given~\cite{virieux2009overview}, because the nonlinearity involved in FWI is strong. It's application is limited in some configurations owing to a number of problems, such as computational inefficiency and human intervention.

Recently, deep-learning~(DL) techniques~\cite{lecun2015deep}, in particular, convolutional neural network~(CNN)~\cite{lecun2010convolutional}, have achieved the state-of-the-art performance in a variety of applications. They surpass the conventional approaches in many research fields of inverse problem, for instance, image reconstruction~\cite{mccann2017convolutional,yang2021robust}, x-ray computed tomography~\cite{jin2017deep}, and optical diffraction tomography~\cite{sun2018efficient,yang2020deep}.  To some extent, this development results from the wide availability of domain-specific languages~(DSL) for DL, such as Tensorflow~\cite{abadi2016tensorflow} or PyTorch~\cite{paszke2017automatic}, used in academia and industry.

Generative adversarial networks~(GANs)~\cite{goodfellow2014generative} are powerful DL models capable of generating realistic images, videos, and voice. Rooted in game theory, one neural network, termed as discriminator, attempts to judge whether the information is real or fake, the other neural network, referred as generator, strives to produce data that the discriminator assumes is real. Wasserstein GAN~(WGAN)~\cite{arjovsky2017wasserstein} is an extension to the GAN, with it's usage not only the training stability is improved but also the loss function is designed to correlate with the quality of generated images. This conceptual shift is motivated mathematically by an argument that training the GAN that generator should seek to minimize the distance between the data distribution observed in the training dataset and the distribution observed in the generated examples. The argument contrasts the Earth-Mover distance, also called Wasserstein-1 distance~\cite{vallender1974calculation}, to achieve this goal.

In this paper, we propose a promising alternative of physics-informed training-free frameworks for two-dimensional~(2D) FWI, termed as FWIGAN. Inspired by~\cite{gupta2021cryogan}, we integrated a physics generator with adversarial learning to solve the ill-posed nonlinear inverse problem.
To the best of our knowledge, it is the first time that the WGAN with gradient penalty~(WGAN-GP)~\cite{gulrajani2017improved} is adopted for FWI, in which the estimation of velocity model is obtained in a distributional sense. A key feature of the proposed framework is that it can learn the noise distribution included in the noisy observed seismograms so that the ``denoised" inversion results can be well estimated.  In addition, FWIGAN is an appealing solution to the peculiar challenges of FWI, such as the usage of an undesired initial model or contaminated measurements. Beyond that, we further illustrate the prospective ability of FWIGAN to simultaneously estimate an uncomplicated source wavelet in FWI workflow.
To summarise, the main contributions of our work are as follows.

\begin{enumerate}
    \item We propose an unsupervised learning framework to handle the inversion task of 2D velocity model from observed seismograms. This paradigm needs no labelled training dataset nor pretraining of networks, and is flexible on account of automatic  differentiation~(AD) of DSL to estimate seismic wavelet synchronously. It requires minimal user interaction such that make this pipeline more available for research.
    \item Inspired by the work of WGAN, we incorporate the acoustic wave equation and replace the generator by a wave equation. Unlike the traditional FWI optimization, we recover the unknown variables through solving a minmax game by competitive learning. Taking advantages of Wasserstein distance and model-data adjoint-driven fashions, our method has the capability to  produce high quality and physically consistent results in a \textit{likelihood-free} manner.
    \item Thanks to the theoretical characterization of WGAN-GP, the proposed method is able to learn the distribution of additive white Gaussian noise~(AWGN) when the observed data is noisy. Consequently, it bypasses the perturbation  of noise and leads to the ``denoised" inversion results. The substantial improvement makes it more robust and versatile to tackle the realistic applications.
    \item  More specifically, the preliminary results demonstrate that our paradigm is robust and general enough to accommodate either an undesired initial model guess or the contaminated seismic data for the handling of local-minima events.
\end{enumerate}

In Section~\ref{sec:relatedworks}, the related works concerning traditional approaches for FWI, supervised learning approaches for waveform inversion and unsupervised learning approaches for FWI are presented. In Section~\ref{sec:physics}, we introduce the physical model of FWI and formulate the computational problem. In Section~\ref{sec:method}, we describe the details of the proposed method, it consists of review of WGAN-GP, mathematical framework of our method and the architecture of the critic. In Section~\ref{sec:result}, we extensively compare our method against FWI solved by using Adam optimizer with mini-batches data to minimize the least-squares function, for Marmousi, Marmousi2, and Overthrust models with noise-free and noisy data. The quantitative evaluation is illustrated to show the effectiveness and generalization of FWIGAN. In Section~\ref{sec:discuss}, we further discuss the details about the estimation of source wavelet, sensitivity to different initial models, and the possible future works. Finally, we summarize our study in Section~\ref{sec:conclude}.

\section{Related Work}\label{sec:relatedworks}

\subsection{Traditional Approaches for FWI}

FWI with least-squares criterion was originally introduced by Tarantola~\cite{tarantola1984inversion}. It is an effective framework to infer the underground structures based on seismic waveforms, but is known to suffer from the local-minima problem, one typical phenomenon of cycle-skipping, and is sensitive to the lack of low frequencies, data noise, and poor starting model. Over the past decades, many researchers have designed various methods by considering the multiple data components~\cite{bunks1995multiscale} or adding regularization terms~\cite{asnaashari2013regularized} to overcome this limitation.

For the multiscale frameworks, FWI can be implemented in the time domain or frequency domain.  Bunks~\textit{et al.}~\cite{bunks1995multiscale} firstly proposed successive inversion of subset data in time domain by increasing high-frequency, because the low frequencies are less sensitive to artifacts. By contrast, the methods in frequency domain lead to a more natural approach for multiscale schemes by performing successive inversions of increasing frequencies.

With regard to the misfit function, the goal of the least-squares regularization term
is to penalize the roughness of the model, hence defining the so-called Tikhonov regularization~\cite{golub1999tikhonov}. Total variation~(TV) regularization is usually implemented by minimizing the $\ell_1$-norm of the model misfit. It was applied to FWI in~\cite{askan2008full} and the weighted regularization was applied to frequency-domain FWI in~\cite{hu2009simultaneous}. Beyond that, the quadratic Wasserstein metric~($W_2$) from optimal transport has been proven as a better preference for mitigating the trapping of cycle-skipping problem~\cite{engquist2016optimal,yang2018application}, since this metric exhibits advantageous mathematical properties, including convexity w.r.t. time shifts and insensitivity to noise.

\subsection{Supervised Learning Approaches for Waveform Inversion}

As we all know, a fully connected neural network with one hidden layer can approximate any continuous function arbitrarily well, when its hidden layer is large enough~\cite{hornik1991approximation}. In seismic exploration, DL-based works for fault detection~\cite{xiong2018seismic,wang2018automatic,wu2019faultseg3d}, random noise attenuation~\cite{yu2019deep,saad2020deep}, interpretation~\cite{wang2019deep,zhang2020can}~and others~\cite{waldeland2018convolutional,yu2021deep} have been rapidly developed.  To address the 2D velocity model building problem from seismic data, there are many approaches based on DL have been proposed. In~\cite{yang2019deep}, the authors introduced a framework applying an U-Net~\cite{ronneberger2015u} to predict the velocity models from prestack multi-shot seismic data. The numerical experiments compared with FWI showed that the complicated nonlinear mapping from seismic data to velocity model can be well approximated by the convolutional neural network~(CNN). Similar works were described in~\cite{araya2019deep,wu2019inversionnet,wang2020velocity,zhang2020data}, which share the same purpose  by using the universal approximation of different CNN architecture.
Most of the aforementioned methods depend on supervised learning to learn the mapping between the paired input and ground-truth. This pipeline needs a large representative training dataset composed of seismic data and the corresponding ground-truth velocity models, which may not be available in many practical applications. In addition, the results obtained by direct feedforward networks might be inconsistent with the observed measurements due to a lack of equation constraints. As a result, the reliability and generalization of this kind methods can not be guaranteed.

Instead of training a deep neural networks~(NNs) with a lot of data pairs, a relative small-size NN can be learned from additional information obtained by enforcing the physical laws to efficiently solve the physical equations. A new NN-based framework, known as physics-informed neural networks~(PINNs), was proposed by Raissi~\textit{et al.}~\cite{raissi2019physics} to solve the forward and inverse problems involving nonlinear partial differential equations~(PDEs) by integrating available data and mathematical models. Aiming at solving the inversion problem of wave equation,  PINN was encoded with acoustic wave equation~(AWE) to produce an accurate velocity model from analytical solution to an initial boundary value or seismic data~\cite{xu2019physics}. The notable results on 2D synthetic model indicated that PINN is able to infer the velocity distribution through the minimization of composed loss function. In the same spirit, Karimpouli and Tahmasebi addressed 1D wave velocity~(P- and S-wave) and density inversion problem via PINN constrained  by seismic wave equation~\cite{karimpouli2020physics}. By contrast, both elastic and acoustic wave inversion were considered and settled with PINN in~\cite{rashtbehesht2020application}.
The motivation for developing those algorithms is that such prior knowledge or constraints can yield more interpretable  methods based on NN, and can provide accurate and physically consistent predictions. However, the effectiveness of this pipeline still relies on the training set, architecture of the NN, and the components of the loss function.

\subsection{Unsupervised Learning Approaches for FWI}

To bridge the gap between supervised learning approaches and seismic waveform inversion, unsupervised learning which does not need labels is a promising paradigm. In~\cite{wu2019parametric}, a CNN-domain FWI method which uses the concept of deep image prior~(DIP) introduced by Ulyanov~\textit{et al.}~\cite{ulyanov2018deep} was proposed. Under the assumption that the velocity model can be well represented by a generative NN, the authors integrated the CNN with FWI and minimized the misfit between simulated data and observed data by updating the parameters of the CNN.
Similarly, Zhu~\textit{et al.}~\cite{zhu2020integrating} described a framework, named NNFWI, to utilize the NN to generate a physical velocity model, which is then embedded in wave equation to simulate the seismic data. The comparable performance of above works demonstrated that the combination of NNs and PDEs is an appealing solution to build velocity model distribution.

Except for the aforementioned methods, another popular label-free variant to solve time-dependent PDEs is recurrent neural networks~(RNNs)~\cite{mikolov2011extensions}. RNNs are a class of artificial neural networks where connections between nodes form a directed graph along a temporal sequence. This allows it to exhibit temporal dynamic behavior. RNNs can use the previous outputs, which are regarded as current inputs, as well as the internal states (memories) to process variable length sequences of inputs. RNN has also been applied to deal with seismic data reconstruction~\cite{yoon2020seismic}, normal moveout velocity estimation~\cite{fabien2020seismic}, and impedance inversion~\cite{alfarraj2019semisupervised}. Thanks to the AD~\cite{baydin2018automatic} of DSL and the fact that the feedforward computation graph of an RNN seems like the iterative process of time-domain wave propagation, FWI can be well implemented by a suitable RNN. In~\cite{richardson2018seismic}, Richardson indicated that an RNN is able to carry out the time-domain finite difference scheme to simulate the forward modeling of AWE. The numerical experiments showed that the Adam optimizer~\cite{KingmaB14} with mini-batches of seismic data produces quicker convergence than stochastic gradient descent~\cite{bottou2010large} and than L-BFGS-B~\cite{zhu1997algorithm} with entire data. Moreover,  the gradient of cost w.r.t. velocity model obtained by using reverse-mode AD is the same as that computed in the adjoint state method. Yet other similar approaches were proposed in~\cite{sun2020theory,fabien2020seismicv,ren2020physics}. In~\cite{wang2021elastic}, the authors improved upon \cite{richardson2018seismic} to perform inversions simultaneously for multiple parameters in elastic isotropic and anisotropic media.

Those novel frameworks based on untrained NNs showed remarkable performance compared to traditional methods and supervised learning approaches for solving FWI. However, the methods based on DIP require early-stopping rules, otherwise the reconstruction quality will be degraded. In addition, it's hard to understand what priors does the NN capture to represent the final solution. By contrast, the paradigms applying  RNN can be well understood.
It is worthy to mention that both DIP- and RNN-based FWI algorithms need a good starting of initial model and enough capacity of GPU memory to store the necessary parameters. Therefore it's difficult to overcome the instinct challenges of FWI, especially when the observed data is contaminated by noise or the initial model is an unexpected candidate.

\section{Problem Formulation}\label{sec:physics}

In order to mimic the wave propagation, we consider the 2D constant density AWE in time domain, in which only P-wave velocity model is an unknown parameter, as the control equation. Let $\Omega =\{(x,z) \in \mathbb{R}^2: z\geq 0\}$ represents the support of the velocity models, more formally, the AWE can be expressed as
\begin{equation}\label{eq:AWE}
    \frac{1}{v(\mathbf{r})^2}\frac{\partial ^2 u(\mathbf{r},t)}{\partial t^2}=\nabla ^2u(\mathbf{r},t)+q(\mathbf{r},t)\delta(\mathbf{r}-\mathbf{r}_{s}),
\end{equation}
where~$\mathbf{r} \in \Omega$~denotes the position, $u: [\Omega \times t] \mapsto \mathbb{R}$~is the wavefield amplitude, $v: \Omega \mapsto \mathbb{R}$~indicates the P-wave velocity distribution, $q$~is the source term with the position constraint~$\mathbf{r}_s$, and~$t$~is the time coordinate. $\nabla^2$~denotes the Laplace operator and~$\delta$~is the Dirac delta function.

For a given velocity model~$\mathbf{v}$ and source term~$q(\mathbf{r}_s,t)$,  the simulated data can be expressed as
$u_s(\mathbf{r}_g,t;\mathbf{v})=Ru(\mathbf{r}_g,t)$, where $R$ denotes the extraction operator that outputs the wavefield $u(\mathbf{r},t)$  at the receiver positions $\mathbf{r}_g$. $u(\mathbf{r}_g,t)$ is the solution of Eq~(\ref{eq:AWE}). We denote the true observed data corresponding to the aforementioned parameters is written as~$d_s(\mathbf{r}_g,t)$,
so the classical FWI~($i.e.,$~least-squares FWI)~usually measures the residual defined as the sum of the squared differences between the simulated data and the true observed data. We can invert the velocity models by minimizing the cost function
\begin{equation}\label{eq:FWIloss}
    \min \mathcal{L}(\mathbf{v}) =
    \min\frac{1}{2}\sum\limits_{s=1}^{n_s}\sum\limits_{g=1}^{n_g} \int_{0}^{T}|u_s(\mathbf{r}_g,t;\mathbf{v})-d_s(\mathbf{r}_g,t)|^2 dt,
\end{equation}
where $T$ is the maximum travel time, and $n_s$ and $n_g$ represent the number of sources and receivers, respectively. Using an adjoint-state algorithm~\cite{plessix2006review}, the gradient of Eq~(\ref{eq:FWIloss}) w.r.t. velocity variable $\mathbf{v}$ yiels the following formalism
\begin{equation}
    \frac{\partial\mathcal{L}(\mathbf{v})}{\partial \mathbf{v}}= \sum\limits_{s=1}^{n_s}\int_{0}^{T}\frac{\partial^2 u_s(\mathbf{r}_g,t)}{\partial t^2}u^{\dag}_s(\mathbf{r}_g,t) dt,
\end{equation}
where~$u^{\dag}_s(\mathbf{r}_g,t)$ represents the adjoint wavefield that is the solution of the following adjoint wave equation on the domain $\Omega$ from time $T$ to 0
\begin{equation}\label{eq:ADWE}
    \frac{1}{v(\mathbf{r})^2}\frac{\partial ^2u^{\dag}_s(\mathbf{r},t)}{\partial t^2}=\nabla ^2u^{\dag}_s(\mathbf{r},t)+R^{T}\frac{\partial \mathcal{L}}{\partial u_s},
\end{equation}
The adjoint wavefield is propagating along a reversal time direction.  Then the velocity model can be updated by it's gradient information iteratively
\begin{equation}
    \mathbf{v}_{k+1} = \mathbf{v}_k -\tau_k M_k \frac{\partial\mathcal{L}(\mathbf{v})}{\partial \mathbf{v}}.
\end{equation}
where $k$ refers to the number of iteration, $\tau_k$ denotes the optimal step length of $k$th iteration, and $M_k$ is the modification of the gradient like conjugate gradient.

\section{Methodology}\label{sec:method}
\begin{figure*}[tp]
    \centering
    \hspace{-0.1cm}
    \includegraphics[width=0.98\textwidth]{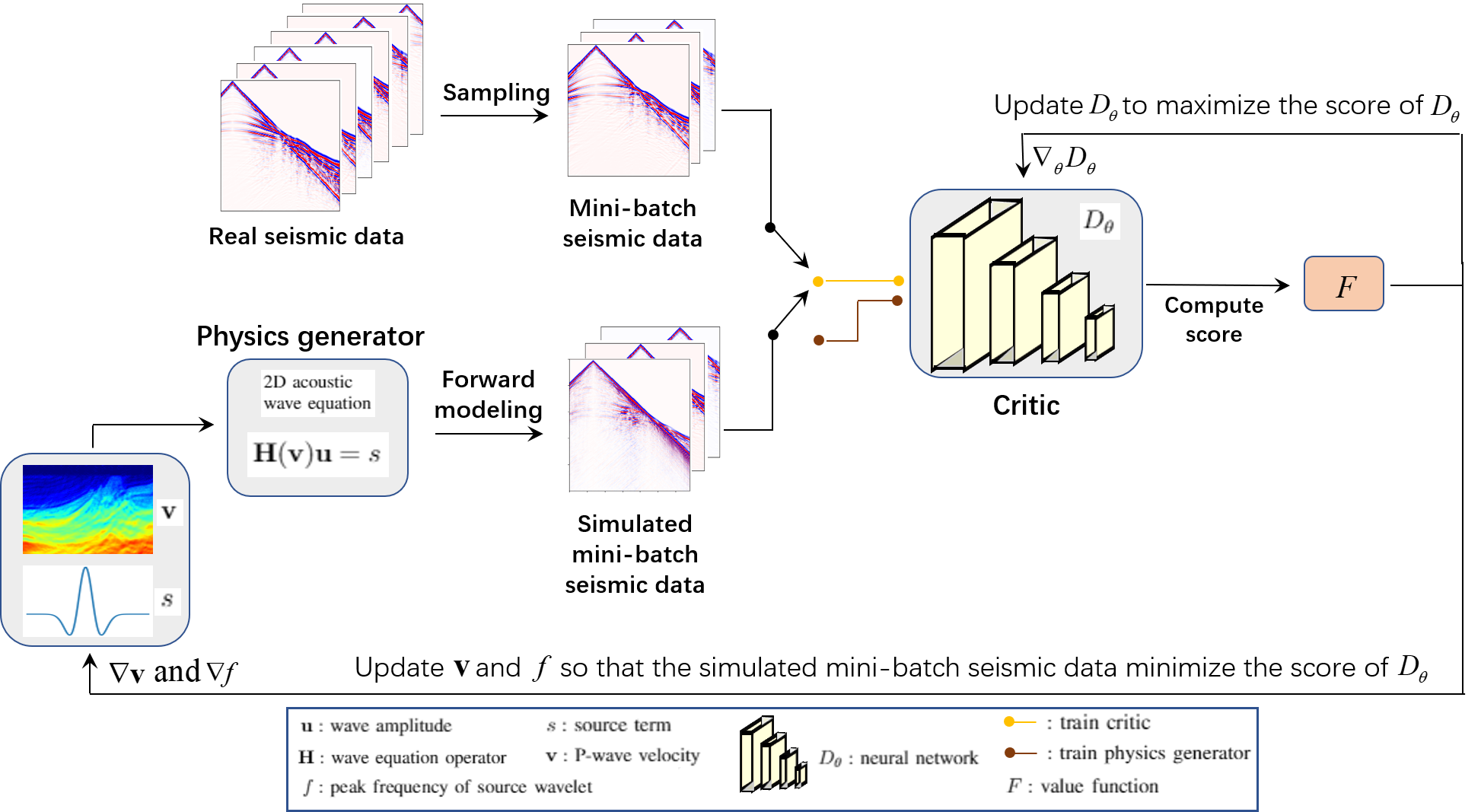}
    \caption{Schematic workflow of the proposed method for seismic waveform inversion with noise-free seismic data.}
    \label{fig:workflow}
\end{figure*}

Generally, the approaches to solve an ill-posed inverse problems can be classified into two categories: model-based and learning-based methods. The first type frameworks aim at obtaining the acceptable solutions with some model-constraint optimization algorithms, while the second kind pipelines usually train a network or atoms through an optimization of a loss function based on the paired input and ground-truth. The former is flexible and general to address diverse inverse problems but is time-consuming due to the sophisticated algorithms. In contrast, the later is restricted by specific tasks while is capable of handling challenging issues and fast for prediction. Inspired by~\cite{gupta2021cryogan}, we combine these two ways which leverage their respective merits to handle the nonlinear inversion task of FWI.

\subsection{Review of WGAN-GP}

According to the work of WGAN, the critic, whose role is similar as that of discriminator of GAN but is not trained to classify, should be in the set of 1-Lipschitz functions. To enforce this constraint,   Arjovsky~\textit{et al.}~\cite{arjovsky2017wasserstein} proposed to clip the weights of the critic to lie in a compact space $[-c,c]$. However the weights clipping strategy leads to optimization difficulties thus even the deep WAGN critics fail to converge sometimes. In~\cite{gulrajani2017improved}, the authors introduced an alternative way, gradient-penalty WGAN, to solve this issue. Consider the fact that a differentiable function is 1-Lipschitz if and only if it has gradients with norm at most 1 everywhere, Gulrajan~\textit{et al.}~\cite{gulrajani2017improved} introduced a novel strategy with gradient penalty, the loss of WGAN-GP is denoted by
\begin{equation}
L = \mathbb{E}_{\tilde{\mathbf{x}}\sim\mathbb{P}_g}[D(\tilde{\mathbf{x}})] - \mathbb{E}_{\mathbf{x}\sim\mathbb{P}_r}[D(\mathbf{x})] + \lambda \mathbb{E}_{\hat{\mathbf{x}}\sim\mathbb{P}_{\hat{\mathbf{x}}}}[(\|\nabla_{\hat{\mathbf{x}}}D(\hat{\mathbf{x}})\|_2-1)^2].
\end{equation}
where $\mathbb{P}_r$  and $\mathbb{P}_g$ are the distribution of true data and generated samples, respectively. $\hat{\mathbf{x}}$ is the random sample satisfies the distribution $\mathbb{P}_{\hat{\mathbf{x}}}$. $D$ denotes the set of 1-Lipschitz functions and $\mathbb{E}$ represents the expectation. $\lambda \in \mathbb{R}_{+}$ is a suitable penalty weight.

By contrast, WGAN-GP makes the training more stable and easier to train, so that we can use a more complex network to implement the tasks.

\subsection{Mathematical Framework of FWIGAN}

Let the region of interest $\Omega$ be discretized into $M = x \times z$ pixels, a general formulation of Eq~(\ref{eq:AWE})~reads as
\begin{equation}\label{eq:forward}
   \mathbf{H}(\mathbf{v})=\mathbf{u}.
\end{equation}
where $\mathbf{v}\in \mathbb{R}^M$ denotes the discretized velocity model, $\mathbf{u}\in \mathbb{R}^V$ is the simulated seismic data of size $V=ns\times T \times ng$,
and $\mathbf{H}: \mathbb{R}^M \mapsto \mathbb{R}^V$~represents the wave equation operator. In addition, $\mathbf{d}\in \mathbb{R}^V$ is the noise-free observed data.

Let $\mathcal{X}$ be a compact metric set and Prob($\mathcal{X}$) denote the space of probability measures defined on $\mathcal{X}$. $ \mathbb{P}_g, \mathbb{P}_r \in$ Prob($\mathcal{X}$) are two distributions. The  Earth-Mover distance or Wasserstein-1 is given by
\begin{equation}\label{eq:Wasserstein1}
    W(\mathbb{P}_g, \mathbb{P}_r)=\inf\limits_{\gamma\in \prod(\mathbb{P}_g,\mathbb{P}_r)} \mathbb{E}_{(\mathbf{a}_1,\mathbf{a}_2)\sim \gamma}[\|\mathbf{a}_1-\mathbf{a}_2\|],
\end{equation}
where $\prod(\mathbb{P}_g,\mathbb{P}_r)$ is the set of all adjoint distribution whose marginals are $\mathbb{P}_g$ and $\mathbb{P}_r$, respectively. Using the Kantorovich-Rubinstein duality~\cite{villani2009optimal}, we can simplify the calculation to
\begin{equation}\label{eq:Wasserstein2}
    W(\mathbb{P}_g, \mathbb{P}_r)=\sup\limits_{\|w\|\leq 1} \mathbb{E}_{\mathbf{a}\sim \mathbb{P}_r}[w(\mathbf{a})]- \mathbb{E}_{\mathbf{a}\sim \mathbb{P}_g}[w(\mathbf{a})],
\end{equation}
where  the supremum is over all the 1-Lipschitz functions $w:\mathcal{X}\rightarrow \mathbb{R}$.

The goal of our task is to estimate the velocity distribution $\mathbf{v}_\mathrm{rec}$, and source peak frequency $f_\mathrm{rec}$ if needed, whose simulated seismic data computed according to Eq~(\ref{eq:forward}) is consistent with the observed data of the true velocity distribution $\mathbf{v}_{\mathrm{true}}$. We assume the conditional probability density function of seismic data $\mathbf{u}$ given a velocity model $\mathbf{v}$ is defined as $\mathbb{P}(\mathbf{u}|\mathbf{v})$. Therefore the inversion task can be formulated as the optimization problem
\begin{equation}\label{eq:obj1}
    \hat{\mathbf{v}}= \arg \min\limits_{\mathbf{v}}\mathcal{M}(\mathbb{P}(\mathbf{u}|\mathbf{v}),\mathbb{P}(\mathbf{d}|\mathbf{v}_\mathrm{true})),
\end{equation}
where $\mathcal{M}$ denotes some distance metric between two distributions.
By using the Wasserstein-1 distance~($i.e.,$~Eq~(\ref{eq:Wasserstein1})), the minimization of Eq~(\ref{eq:obj1}) translates as
\begin{equation}\label{eq:obj2}
    \hat{\mathbf{v}} = \arg \min\limits_{\mathbf{v}}\inf\limits_{\gamma\in \prod(\mathbb{P}_\mathrm{rec},\mathbb{P}_\mathrm{true})} \mathbb{E}_{(\mathbf{u}_1,\mathbf{d}_1)\sim \gamma}[\|\mathbf{u}_1-\mathbf{d}_1\|],
\end{equation}
where $\mathbb{P}_\mathrm{rec}=\mathbb{P}(\mathbf{u}|\mathbf{v})$ and $\mathbb{P}_\mathrm{true}=\mathbb{P}(\mathbf{d}|\mathbf{v}_\mathrm{true})$.
Note that it is necessary to do the normalization to turn seismic signals
into probability measures such that the theory of optimal transport applies. We adopt the strategy of data normalization via a linear transformation and rescaling proposed in~\cite{yang2018application} as follows
\begin{equation}\label{eq:noemalize}
    P(\mathbf{u})=\frac{\mathbf{u}+c}{\sum(\mathbf{u}+c)},
\end{equation}
where $c$ is a constant such that $\mathbf{u}+c>0$. In practice, we prefer to choose $c$ approximate 1.1 times absolute value of the minimum value of the observed data. This prepsocessing can ensure the positivity of both simulated and observed seismic signals. After that, we rescale all signals to unit mass 1. The normalized data will be the candidate fed into the critic.

\begin{algorithm}[t]
\caption{FWIGAN\label{algo:FWIGAN}}
\textbf{Input:}~$\mathbf{v}_0:$~initial velocity model, $\mathbf{H}:$~acoustic wave equation operator, $n_s:$~number of source, $\{\mathbf{d}\}_{i=1}^{n_s}:$~real common-source seismograms, $E:$~number of epochs, $n_\mathrm{critic}:$~number of iterations of training critic per updating velocity model, $B:$~batch size, $\lambda:$~penalty weight, $D_{\mathbf{\theta}_0}:$~initialized critic, $lr_v:$~learning rate for updating velocity model,  $lr_c:$~learning rate for training critic.
\begin{algorithmic}[1]
	\FOR {$e=1,\ldots,E$}
	\FOR {$b=1,\ldots,B$}
	\FOR {$n=1,\ldots,n_\mathrm{critic}$}
	\STATE sample data $\{\mathbf{d}^1,\ldots,\mathbf{d}^A\}$ with $A=n_s/B$ from real dataset $\{\mathbf{d}\}_{i=1}^{n_s}$
	\STATE generate the corresponding simulated data $\{\mathbf{u}^1,\ldots,\mathbf{u}^A\}$ from current velocity model using Eq~(\ref{eq:forward})
	\STATE uniformly sample $\mu_1,\ldots,\mu_A$ from $U(0,1)$
	\STATE do normalization for $\{\mathbf{d}^i\}_{i=1}^A$ and $\{\mathbf{u}^i\}_{i=1}^A$ using Eq~(\ref{eq:noemalize})
	\STATE update the network parameters $\mathbf{\theta}$ according to $\nabla_{\mathbf{\theta}}\mathcal{L}(\mathbf{v},D_\mathbf{\theta})$  of Eq~(\ref{eq:obj4}) using Adam optimizer with learning rate $lr_c$
 	\ENDFOR
 	\STATE generate simulated data $\{\mathbf{u}^1,\ldots,\mathbf{u}^A\}$ using Eq~(\ref{eq:forward})
 	\STATE update $\mathbf{v}$ according to $\nabla_{\mathbf{v}}\mathcal{L}(\mathbf{v},D_\mathbf{\theta})$ of Eq~(\ref{eq:obj4}) using Adam optimizer with learning rate $lr_v$
 	\ENDFOR
 	\ENDFOR
\RETURN  final inverted velocity model $\hat{\mathbf{v}}$
\end{algorithmic}
\end{algorithm}

According to the theory of WGAN-GP, we can estimate the velocity model by optimizing the cost function
\begin{equation}\label{eq:obj3}
\hat{\mathbf{v}} = \arg \min\limits_{\mathbf{v}}\max\limits_{D_{\mathbf{\theta}}} \mathbb{E}_{\mathbf{d}\sim\mathbb{P}_\mathrm{true}}[D_{\mathbf{\theta}}(P(\mathbf{d}))] - \mathbb{E}_{\mathbf{u}\sim\mathbb{P}_\mathrm{rec}}[D_{\mathbf{\theta}}(P(\mathbf{u}))]
- \lambda \mathbb{E}_{\hat{\mathbf{u}}\sim\mathbb{P}_\mathrm{int}}[(\|\nabla_{\hat{\mathbf{u}}}D_{\mathbf{\theta}}(\hat{\mathbf{u}})\|_2-1)^2],
\end{equation}
where $D_{\mathbf{\theta}}$ denotes the critic~($i.e.,$ a neural network) with parameters $\mathbf{\theta}$. The Lipschitz constraint $\|D_{\mathbf{\theta}}\|_L<1$ is enforced by penalizing the norm of the gradient of $D_{\mathbf{\theta}}$ output w.r.t. its input. $\mathbb{P}_\mathrm{int}$ is implicitly defined as sampling uniformly along straight lines between pairs of points sampled from the real data distribution $\mathbb{P}_\mathrm{true}$
and the generated data distribution $\mathbb{P}_\mathrm{rec}$, $i.e., \hat{\mathbf{u}}=\mathbf{\mu} P(\mathbf{d})+(1-\mathbf{\mu})P(\mathbf{u})$ with $\mathbf{\mu}$ uniformly sampled between 0 and 1. This is motivated by the fact that the optimal critic contains straight lines with gradient norm 1 connecting
coupled points from $\mathbb{P}_\mathrm{true}$ and $\mathbb{P}_\mathrm{rec}$.

By replacing the expectation by the empirical values, the min-max optimization problem of Eq~(\ref{eq:obj3}) leads to the following final formulation
\begin{equation}\label{eq:obj4}
\mathcal{L}(\mathbf{v},D_{\mathbf{\theta}})=\sum\limits_{n\in N}[D_{\mathbf{\theta}}(P(\mathbf{d}^n))] - \sum\limits_{n\in N}[D_{\mathbf{\theta}}(P(\mathbf{H}(\mathbf{v})^n))]
-\lambda\sum\limits_{n\in N}[(\|\nabla_{\hat{\mathbf{u}}}D_{\mathbf{\theta}}(\hat{\mathbf{u}}^n)\|_2-1)^2].
\end{equation}
Here, $N$ denotes either entire common-shot seismograms $N_{\mathrm{all}}=\{1,\ldots,n_s\}$ or mini-batches sampled from full $A\subseteq N_{\mathrm{all}}$. $P(\mathbf{d}^n)$ is the normalized real observed data according to Eq~(\ref{eq:noemalize}) and $P(\mathbf{H}(\mathbf{v})^n)$ is the normalized simulated data obtained from the current velocity model and source peak frequency by physics generator.

The schematic diagram of FWIGAN is shown in Figure~\ref{fig:workflow}. The workflow resembles that of original WGAN,  with a mutation that the AWE operator, named as physics generator, substitute for the generative neural network.  FWIGAN is driven by the competitive ``training" of two sides: the critic discriminates the difference between the observed data $\mathbf{d}$ and the generated data $\mathbf{u}$, yet the physics generator aims to physically simulate seismic data from current velocity model. Gradients from the critic backpropagates to $\mathbf{v}$ to update it directly at each optimization step. Unlike the traditional FWI algorithms, FWIGAN attempts to reconstruct the variable $\mathbf{v}$ by playing a minmax game to make the simulated data most closely match the real data. Thanks to the adversarial learning strategy, FWIGAN is able to produce high-resolution and physics-constraint predictions. The process is described in Algorithm~\ref{algo:FWIGAN}.

\begin{figure*}[tp]
    \centering
    \includegraphics[width=0.8\textwidth]{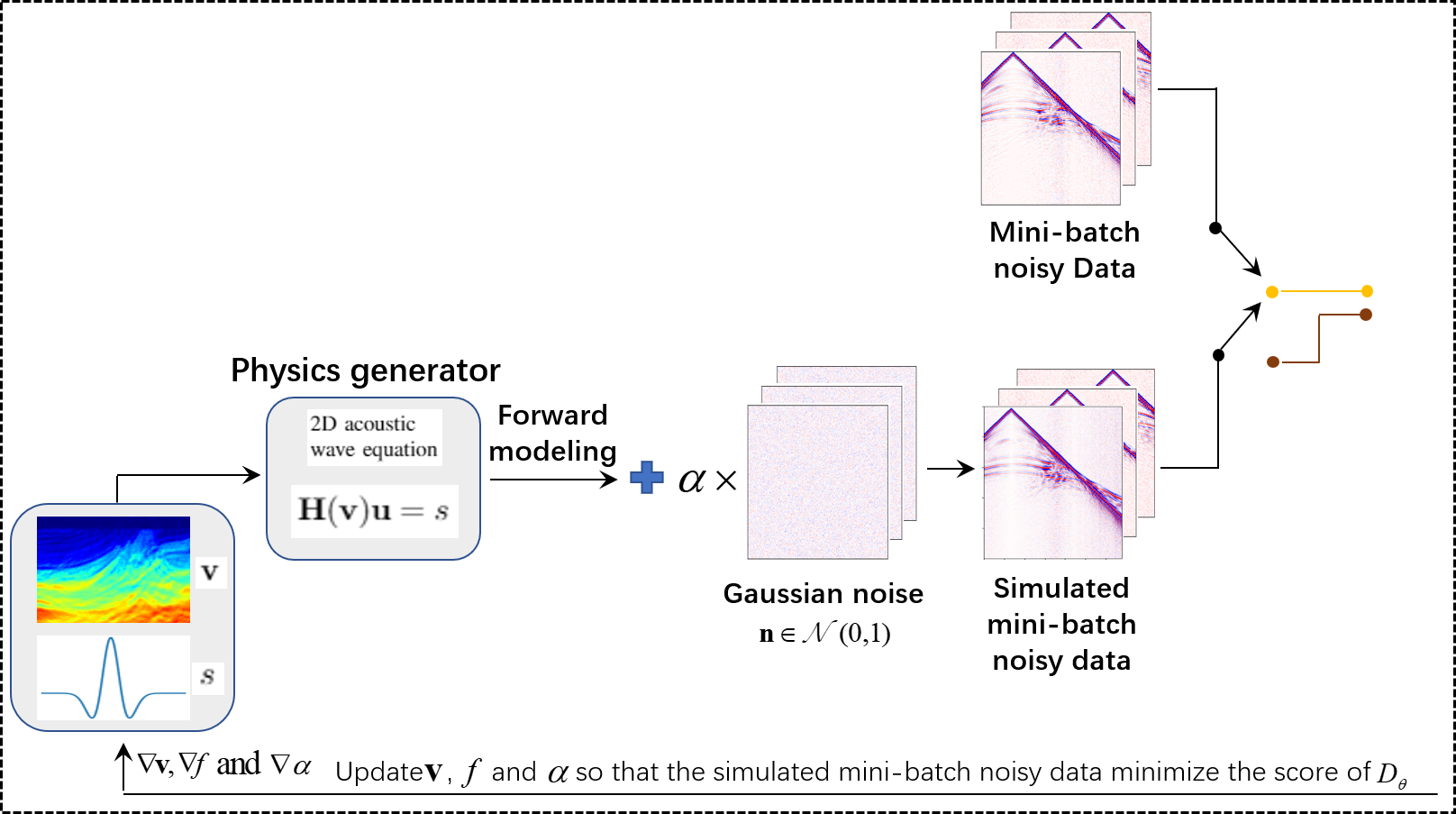}
    \caption{Modified section of the proposed method for seismic waveform inversion with noisy seismic data.}
    \label{fig:noisyworkflow}
\end{figure*}

Note that the aforementioned observed data $\mathbf{d}$ is assumed noise-free, which means it is generated from the ground-truth velocity model $\mathbf{v}_\mathrm{true}$ by using the same forward modeling algorithm of Eq~(\ref{eq:forward}) as simulated data. As with any FWI implementation, you are unlikely to get such nice results on real data. Therefore, we further explore the feasibility of FWIGAN for velocity inversion with noisy seismic data.
Under assumption that the noise exists in the noisy observed data is AWGN that satisfy the distribution $\mathcal{N}(0,\sigma^2)$, so the noisy seismograms can be expressed as $\mathbf{d}_\mathrm{noisy}=\mathbf{d}+\mathbf{n}_\sigma$. Accordingly Eq~(\ref{eq:forward}) can be reformulated as
\begin{equation}\label{eq:forward1}
   \mathbf{H}(\mathbf{v})+\mathbf{n}=\mathbf{u}_\mathrm{noisy},
\end{equation}
where $\mathbf{n}$ denotes the AWGN satisfy the distribution $\mathcal{N}(0,\alpha^2)$. For the consistency, we shall hence use $\mathbf{n}_\alpha$ to express the AWGN in Eq~(\ref{eq:forward1}), therefore $\mathbf{u}_\mathrm{noisy} = \mathbf{u}+\mathbf{n}_\alpha$. We denote the conditional probability density function for the noisy scenario as $\mathbb{P}(\mathbf{u}+\mathbf{n}_\alpha|\mathbf{v})$, the probability distribution of AWGN in noisy observed and simulated data are $\mathbb{P}(\mathbf{n}_\sigma)$ and $\mathbb{P}(\mathbf{n}_\alpha)$, respectively. Then given
$\mathbb{P}(\mathbf{n}_\alpha)=\mathbb{P}(\mathbf{n}_\sigma)$~($i.e.,$~$\alpha=\sigma$), the following holds
\begin{equation}\label{eq:probability}
    \mathbb{P}(\mathbf{u}+\mathbf{n}_\alpha|\mathbf{v})=\mathbb{P}(\mathbf{d}+\mathbf{n}_\sigma|\mathbf{v}_\mathrm{true})\Rightarrow \mathbb{P}(\mathbf{u}|\mathbf{v})=\mathbb{P}(\mathbf{d}|\mathbf{v}_\mathrm{true}),
\end{equation}
Because the AWGN is statistically independent of the seismic signal, then the distribution of the summation of two random variables has the following equality
\begin{equation}
    \mathbb{P}(\mathbf{u}+\mathbf{n}_\alpha|\mathbf{v})=\mathbb{P}(\mathbf{u}|\mathbf{v})\ast \mathbb{P}(\mathbf{n}_\alpha),
\end{equation}
That is, the probability distribution of the sum of two or more independent random variables is equal to the convolution (denoted by $\ast$) of their individual distributions~\cite{hogg2005introduction}. This implies that

\begin{equation}
    \mathcal{F}\{\mathbb{P}(\mathbf{u}+\mathbf{n}_\alpha|\mathbf{v})\}=\mathcal{F}\{\mathbb{P}(\mathbf{u}|\mathbf{v})\}\mathcal{F}\{\mathbb{P}(\mathbf{n}_\alpha)\},
\end{equation}
where $\mathcal{F}$ is the Fourier transform. We assume $\mathbb{P}(\mathbf{n}_\sigma)$ has  a full support in Fourier domain, $i.e., \{\mathcal{F}\mathbb{P}(\mathbf{n}_\sigma)\}(\omega) \neq 0, \forall \omega$, thus we can obtain
\begin{equation}
    \mathbb{P}(\mathbf{u}|\mathbf{v})=\mathcal{F}^{-1}\Big\{\frac{\mathcal{F}\{\mathbb{P}(\mathbf{u}+\mathbf{n}_\alpha|\mathbf{v})\}}{\mathcal{F}\{\mathbb{P}(\mathbf{n}_\alpha)\}}\Big\}.
\end{equation}

From this, it is easy to see that Eq~(\ref{eq:probability}) holds when $\alpha=\sigma$.

Taking advantages of AD, we can also parameterize the  noise level and estimate its value during the training process. Although the critic distinguishes the difference between the noisy real data and the noisy generated data, the framework of FWIGAN is still  able to capture the ``denoised" velocity distribution if $\alpha$ approximates to $\sigma$. The modified scheme is shown in Figure~\ref{fig:noisyworkflow}.
It is worthy to note that the variable $\alpha$ would be added to the unknown parameters when we also restore the noise level. Hence, the loss function we optimize corresponds to the noisy configuration is
\begin{equation}\label{eq:obj5}
\mathcal{L}(\mathbf{v},D_{\mathbf{\theta}},\alpha)=
\sum\limits_{n\in N}[D_{\mathbf{\theta}}(P(\mathbf{d}^n+\mathbf{n}_\sigma))]-
 \sum\limits_{n\in N}[D_{\mathbf{\theta}}(P(\mathbf{H}(\mathbf{v})^n+\mathbf{n}_\alpha))]-
\lambda \sum\limits_{n\in N}[(\|\nabla_{\tilde{\mathbf{u}}}D_{\mathbf{\theta}}(\tilde{\mathbf{u}}^n)\|_2-1)^2].
\end{equation}
where $\tilde{\mathbf{u}}$ is the uniformed sampling corresponding to the noisy observed and simulated data.
\begin{figure*}[tp]
    \centering
    \includegraphics[width=1.\textwidth]{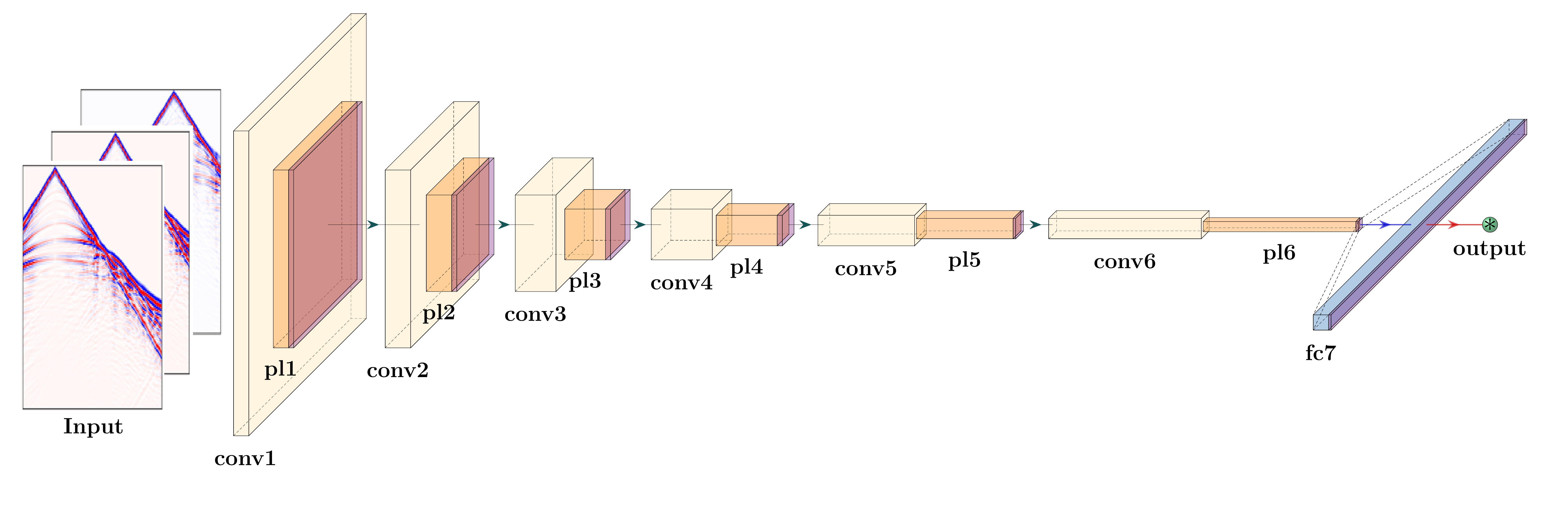}
    \caption{Architecture of the critic used for seismic waveform inversion. Multiple shot gathers are input to the critic simultaneously, which is substantially different from conventional FWI. The network is composed by repeated convolutional layer~(yellow box), max-pooling layer~(orange box), leaky ReLU operation~(purple box) and fully connected layer~(blue box). The final output is a scalar.}
    \label{fig:architecture}
\end{figure*}

\subsection{The Critic of FWIGAN}

Indeed, the critic of FWIGAN is similar to those used in the standard WGANs. It plays a role which is close to the discriminator of GANs, but without the sigmoid function and outputs a scalar score rather than a probability. This score can be interpreted as how real the input images are through measuring the difference between real data and fake data generated by the physics generator.
Driven by the adversarial learning scheme, the parameters of critic will be adjusted according to the back-propagation of the loss. Meanwhile, it will provide the feedback information to update the velocity model to fit the fake data with real data well.

We design a CNN based on that of WGAN, the network consists of repeated applications of six convolutional blocks and two fully connected layers.
The components of one block are
\begin{enumerate}
    \item A ($3 \times 3$)  2D convolutional layer with stride ($1 \times 1$)
    \item A ($2 \times 2$) max-pooling layer with stride ($2 \times 2$)
    \item A leaky ReLU function with negative slope of 0.1
\end{enumerate}
The max-pooling layer leads to downsampling such that the size of the feature map is halved. The feature map of last convolutional block is flatten to a vector, then fed into a fully connected layer with neurons of 2000 followed by the leaky ReLU function. The final output is obtained in the form of a scalar. Note that the batch normalization is omitted to safeguard the gradient penalty strategy. Furthermore, the number of channels in first convolutional block is 32, then is doubled in the next block successively. We set the input channels is equal to the number of the shots of mini-batch seismic data. The architecture of the critic adopted for FWI is described in Figure~\ref{fig:architecture}.

\section{Numerical Experiments}\label{sec:result}

In this section, we present experiments that validate our proposed method~(FWIGAN) on 2D synthetic models. We compare FWIGAN against the state-of-the-art FWI method which is implemented by using Adam optimizer to minimize the least-squares function. Since source excitation is generally unknown and should be considered as an estimated parameter in seismic inversion like time-domain FWI, we did the tasks of velocity inversion and source wavelet estimation ($i.e.,$ reconstruction of dominant frequency of wavelet)  for three models simultaneously. In the inversion process, we avoid the use of techniques such as
adding regularization.
The optimization is performed on a desktop workstation (Nvidia GeForce GPU, Ubuntu operating system) and implemented on PyTorch, especially the toolbox named Deepwave\footnote{The link for Deepwave is available from \url{https://github.com/ar4/deepwave}}~\cite{richardson2018seismic}. It provides wave propagation modules and allows to automatically reconstruct the unknown variables owing to the chain operations and back-propagation.
\begin{figure*}[tp]
    \centering
    \includegraphics[width=0.8\textwidth]{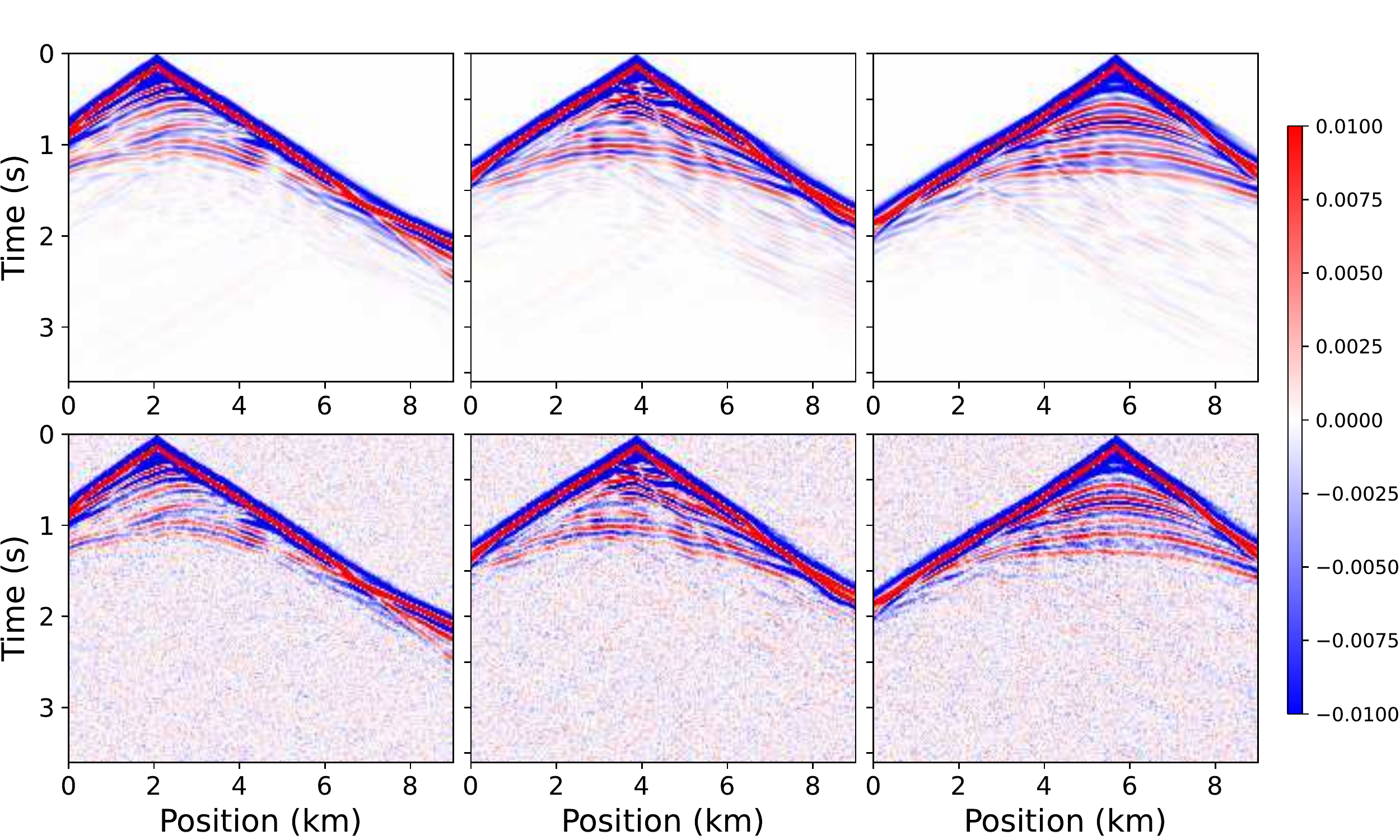}
    \caption{Three common-source gathers generated from Overthrust model by using finite difference scheme. For visualization, the data is normalized with the division of maximum absolute value of clean data. First row: noise-free seismic data. Second row: corresponding noisy data that the SNR is 10dB.}
    \label{fig:data}
\end{figure*}

\subsection{Data Preparation}

To illustrate the availability of our proposed method, we test the inversion performance for three well-known 2D synthetic models which are suitable for a wide variety of geophysical research, that is Marmousi model~\cite{versteeg1994marmousi}, Marmousi2 model~\cite{martin2006marmousi2}, and Overthrust model~\cite{aminzadeh1994seg}. With the current limitation of computational equipment, we downsampled the model size and aim to restore the rescaled models.

\textbf{Marmousi Model:}
One of the most famous and standard acoustic velocity models in exploration geophysics,  was created by the French Institute of Petroleum in 1988 and its geometry is designed based on seismic profiles in the Cuanza Basin, Angola.
We resampled the P-wave velocity model, subsequently, the size is $(x\times z)=(100\times 310)$ with a spatial grid increment of 0.03 km. The velocity ranges from 1472 m/s to 5772 m/s. The true model is shown in Figure~\ref{fig:mar1v}.

\textbf{Marmousi2 Model:}
An updated version based  on the original Marmousi structure, but is extended in width and depth, and is made fully elastic.  The original Marmousi model is close to the center of Marmousi2. By contrast, Marmousi2 is better to simulate long-offset acquisition in a deepwater setting. In our work we use only the P-wave velocity for data simulation and inversion. The downsampled model size is $(x \times z)=(130 \times 340)$ and its velocity is ranging from 1140 m/s to 4700 m/s. The first 16 grid in depth represent the structure of  water with a velocity value of 1500 m/s.
We set the spacial increment as 0.03 km and the true model is illustrated in Figure~\ref{fig:mar2v}.

\textbf{Overthrust Model:}
A 3D geological model that was created as part of a multi-phase collaboration between the SEG and EAGE, we take 2D slice of original model and attempt to reconstruct it. The model is overlain by a flat sedimentary layer underneath the sea bed. It represents a varying degree of complexity, with a central thrust faulted anticline, and an external monocline and flat zone. The resampled model size is $(x\times z)=(90\times 300)$ with a spacial increments of 0.03 km. The velocity value is in the range of 2360 m/s - 6000 m/s. The true model is present in Figure~\ref{fig:overv}.

\subsection{Parameter Setting}

\textbf{Forward Modeling:}
In order to simulate the wave propagation constraint by AWE, a regular grid and finite difference scheme~\cite{strikwerda2004finite}, with second-order accuracy in time domain and forth-order accuracy in space domain, were applied for the forward modeling of FWI and FWIGAN.
To prevent unwanted reflections from the edges of the simulation domain, we used a perfectly matched layer~(PML)~\cite{komatitsch2003perfectly} absorbing boundary condition. A Ricker wavelet with a peak frequency of 7 Hz was chosen as the true source wavelet. The time interval for forward simulation was 3 ms. In terms of three different models,  the total recording time was set as 6s, 7.5s, and 3.6s, and thirty sources were equally placed on the surface ($i.e., z=0$) with a horizontal interval of 0.3 km, 0.329 km, and 0.29 km,  respectively. Accordingly, receivers were evenly arranged on the surface with a interval of 0.29 km. The number of receivers is equal to the width of the model. Three normalized common-source gathers generated from Overthrust model are shown in Figure~\ref{fig:data}.

In addition, to make the observation more realistic, we added AWGN $\mathbf{n}$ to the noise-free seismic data $\mathbf{d}$. The noise was added such that the input SNR = 10dB, where SNR$(\mathbf{d}+\mathbf{n},\mathbf{d})=20\log_{10}(\|\mathbf{d}\|_2/\|\mathbf{n}\|_2)$. Three examples are also displayed in Figure~\ref{fig:data}. Note that the seismograms generated from true velocity models by using aforementioned forward modeling algorithm are regraded as the true observed data for inversion task. Same simulation scheme is adopted to act as the physics generator to produce the simulated data. Similar for the noisy case.

\begin{figure*}[tp]
    \centering
    \includegraphics[width=0.9\textwidth]{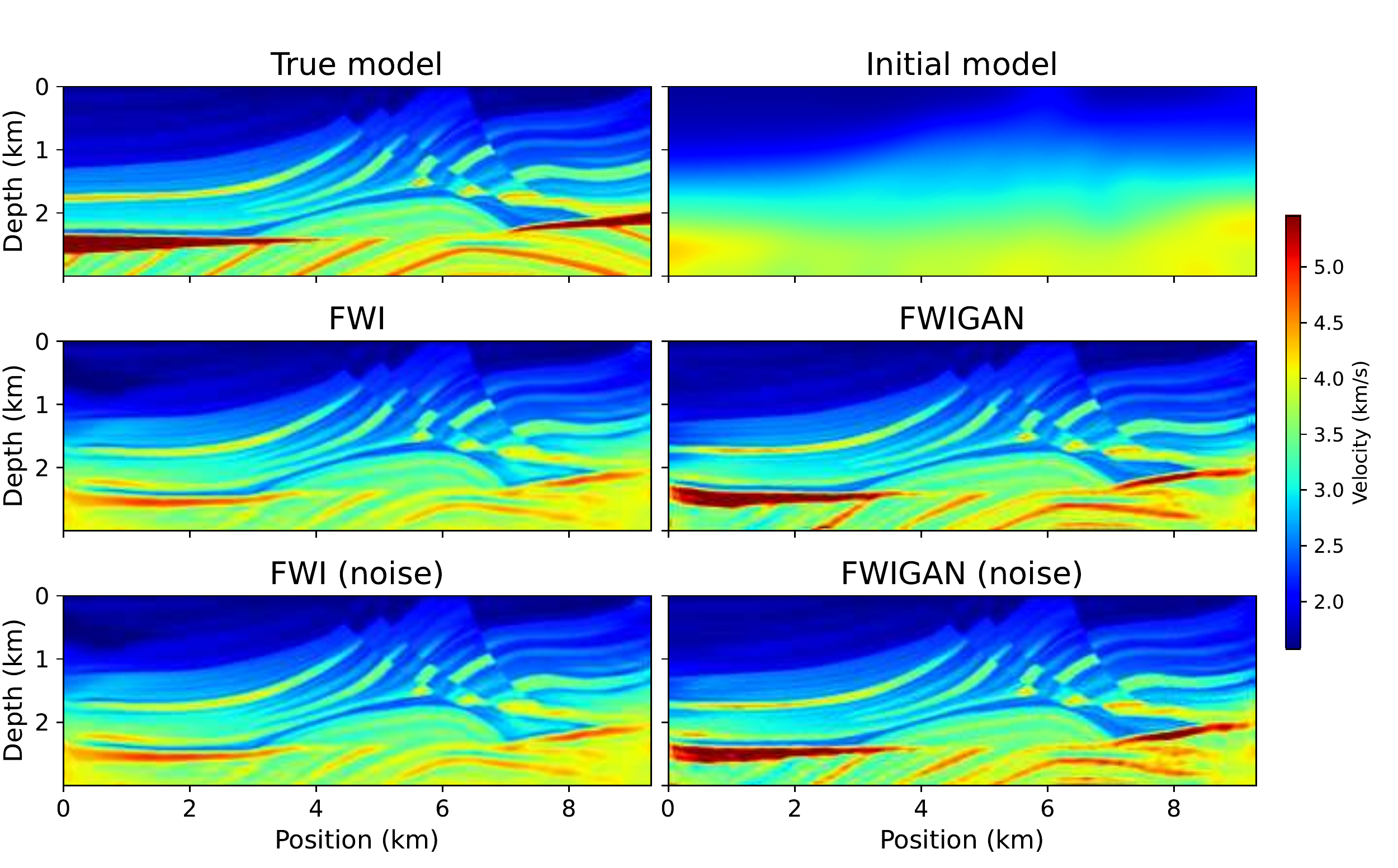}
    \caption{Comparable inversion results of Marmousi model. Frist row: true P-wave velocity model and initial guess. Second row: inverted results by FWI and the proposed method~(FWIGAN) with noise-free seismic data. Third row: inverted results by FWI and the proposed method~(FWIGAN) with noisy seismic data. The model size is $100 \times 310$ with a spacial interval of 0.03 km.}
    \label{fig:mar1v}
\end{figure*}

\textbf{FWI with $\ell_2$ norm:}
Be benefit from the automatic differentiation of PyTorch, it's easy to automatically adjust the parameters when the differentiated functions are given. Seismic FWI can be well carried out by using Adam optimizer~$(\beta_1=0.5, \beta_2=0.9, \epsilon=10^{-8})$ with mini-batches data. The initial model guess is obtained by Gaussian
smoothed function with the true model~(see Figure~\ref{fig:mar1v}-Figure~\ref{fig:overv}). According to the guidelines provided by the researchers, we did normalization processing that consists of division of maximum absolute value and rescaling for both observed and simulated mini-batches data. The batch size was set as 1 and the learning rate we used for updating velocity models was 50. The algorithm was run for 800 epochs while the learning rate was decreased by 0.5 at 100th and 200th epoch.  The minimum value constraint was adopted to improve the inversion performance.

\subsection{Optimization Strategy of FWIGAN}

For the implementation of FWIGAN, we did data normalization according to Eq~(\ref{eq:noemalize}) for each mini-batches data during the training process. The initial model guess and true observed data are as same as those of FWI.
The optimization was realize by using the same Adam optimizer. We selected the learning rate of 5 for updating the velocity model and $10^{-3}$ for changing the parameters of critic.  The weight for the gradient penalty term was kept to $\lambda=10$. Similarly, we decreased the learning rate with a decay of 0.5 at every 100 epochs and the training process was run for 300 epochs totally. The hyperparameters of the critic were initialized to default values by PyTorch. It was trained 6 times~($i.e., n_\mathrm{critic}=6$) for each updating of model. The batch size was set as 5 which means thirty sources were split into 6 batches and each batch had 5 common-shot gathers. In our experiments, the number of the batch size did  not significantly impact the inversion performance once the training converges.

For the back-propagation, the value of the gradients for velocity model were clipped to a maximum value of 10 for noise-free case and $10^3$ for noisy case. Meanwhile, the norm of the gradients for critic were also clipped to a maximum value of $10^3$ for noise-free case and $10^6$ for noisy case.

When the observed data is noisy, we can learn the distribution of noise during the training. The initial guess of SNR was 20dB and the learning rate for updating it was 1. The other optimized strategies were as same as those of velocity model.

\subsection{Quantitative Evaluation}

In order to assess the quality of our proposed method, we quantitatively evaluate the quality of the inverted velocity model $\hat{\mathbf{v}}$ w.r.t. the ground-truth ${\mathbf{v}}$. We compute the structural similarity~(SSIM) defined as
\begin{equation}
    \mathrm{SSIM}(\hat{\mathbf{v}},\mathbf{v}) = \frac{(2\mu_{\mathbf{v}}\mu_{\hat{\mathbf{v}}}+c_{1})(2\sigma_{{\mathbf{v}}}\sigma_{\hat{\mathbf{v}}}+c_{2})}{(\mu_{{\mathbf{v}}}^{2}+\mu_{\hat{\mathbf{v}}}^{2}+c_{1})(\sigma_{{\mathbf{v}}}^{2}+\sigma_{\hat{\mathbf{v}}}^{2}+c_{2})},
    \label{eq:SSIM}
\end{equation}
where $\mu_{{\mathbf{v}}}$, $\mu_{\hat{\mathbf{v}}}$, $\sigma_{{\mathbf{v}}}$, $\sigma_{\hat{\mathbf{v}}}$, and $\sigma_{{\mathbf{v}}\hat{\mathbf{v}}}$ are the local means, standard deviations, and cross-covariance for images ${\mathbf{v}}$, $\hat{\mathbf{v}}$, respectively. The regularization constants {$c_{1}=10^{-4}$ and $c_{2}=9\times 10^{-4}$} evade instabilities over image regions where the local mean or standard deviation is vanishing.

In addition, we compute the relative error between the inverted velocity model $\hat{\mathbf{v}}$ and ground-truth $\mathbf{v}$ defined as
\begin{equation}
    \mathrm{ERROR}(\hat{\mathbf{v}},\mathbf{v}) = \frac{\|\mathbf{v}-\hat{\mathbf{v}}\|_{2}}{\|\mathbf{v}\|_{2}}.
\label{eq:ReError}
\end{equation}
where $\|\cdot\|_2$ denotes the $\ell_2$ norm. The higher SSIM and lower ERROR indicate the better reconstruction.

\subsection{Results of Marmousi Model}

The comparable inversion results by FWI with least-squares minimization and the proposed method~(FWIGAN) are illustrated in Figure~\ref{fig:mar1v}. As expected, with a relatively good starting model, FWI yields a reasonable result (see second row of Figure~\ref{fig:mar1v}). However, it produces underestimated velocity value in the deep area due to a lack of useful information of deep structures contained in the observation.
On the contrary, our framework with adversarial learning strategy faithfully restore the velocity model for this scenerio, especially the deep structures.

To further assess the capacity of FWIGAN to invert the models when the seismic data is contaminated with noise, we proposed to learn the distribution of noise beside the framework simultaneously. The performance is visually present in Figure~\ref{fig:mar1v} (see third row).
The estimated velocity by FWI seems well. By contrast, FWIGAN is more robust to obtain the denoised inversion result. In addition, we summarize in Table~\ref{tb:metric1} and Table~\ref{tb:metric2} the SSIM and ERROR values which indicate our observation. The estimated SNR (see Table~\ref{tb:recSNR}) is 10.36 dB that is close to the true SNR.  It shows that our method is able to handle the realistic applications.

\begin{table}[!tbp]
\setlength{\tabcolsep}{8pt}
\centering
\caption{SSIM and relative error~(ERROR) of the inverted velocity model by using FWI and the proposed method~(FWIGAN) with noise-free seismic data.\label{tb:metric1}}
\arrayrulecolor{black}
\begin{tabular}{c | c | ccc}
\toprule\toprule
Metric & Method  & Marmousi  & Marmousi2 & Overthrust\\
\midrule
\multirow{2}{*}{{SSIM}}
&FWI    &0.7072 &0.6518 &0.7802\\
&FWIGAN   &\textbf{0.8545} &\textbf{0.8099} &\textbf{0.7966}\\
\midrule
\multirow{2}{*}{{ERROR}}
&FWI    &0.1362 &0.1067 &0.0355 \\
&FWIGAN   &\textbf{0.0876} &\textbf{0.0678} &\textbf{0.0269}\\
\bottomrule\bottomrule
\end{tabular}
\end{table}

\begin{table}[!tbp]
\setlength{\tabcolsep}{8pt}
\centering
\caption{SSIM and relative error~(ERROR) of the inverted velocity model by using FWI and the proposed method~(FWIGAN) with noisy seismic data.\label{tb:metric2}}
\arrayrulecolor{black}
\begin{tabular}{c | c | ccc}
\toprule\toprule
Metric & Method  & Marmousi  & Marmousi2 & Overthrust\\
\midrule
\multirow{2}{*}{{SSIM}}
&FWI    &0.6817 &0.6381 &0.7611\\
&FWIGAN   &\textbf{0.8357} &\textbf{0.7324} &\textbf{0.7722}\\
\midrule
\multirow{2}{*}{{ERROR}}
&FWI    &0.1366 &0.1068 &0.0360 \\
&FWIGAN   &\textbf{0.0825} &\textbf{0.0855} &\textbf{0.0310}\\
\bottomrule\bottomrule
\end{tabular}
\end{table}

\begin{table}[!tbp]
\setlength{\tabcolsep}{8pt}
\centering
\caption{Estimated signal-to-noise ratio~(SNR) by using the proposed method~(FWIGAN) when the seismic data is contaminated with additive white Gaussian noise. The true SNR is 10dB.\label{tb:recSNR}}
\arrayrulecolor{black}
\begin{tabular}{ c | ccc}
\toprule\toprule
 & Marmousi  & Marmousi2 & Overthrust\\
\midrule
SNR~(dB) &10.36  &10.11  &10.29\\
\bottomrule\bottomrule
\end{tabular}
\end{table}

\begin{figure*}[tp]
    \centering
    \includegraphics[width=0.9\textwidth]{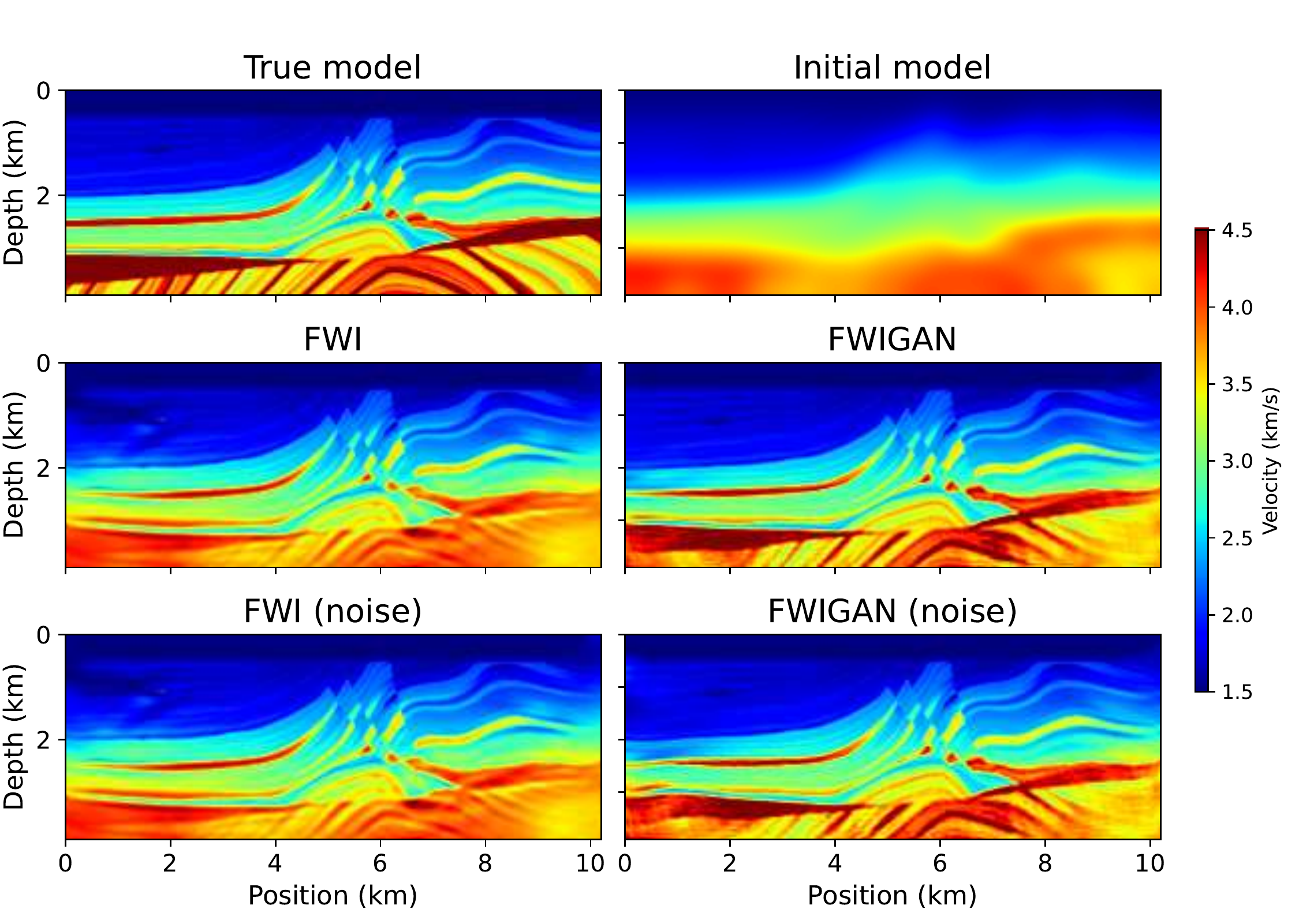}
    \caption{Comparable inversion results of Marmousi2 model. Frist row: true P-wave velocity model and initial guess. Second row: inverted results by FWI and the proposed method~(FWIGAN) with noise-free seismic data. Third row: inverted results by FWI and the proposed method~(FWIGAN) with noisy seismic data. The model size is $130 \times 340$ with a spacial interval of 0.03 km.
    }
    \label{fig:mar2v}
\end{figure*}
\subsection{Results of Marmousi2 Model}

\begin{figure*}[tp]
    \centering
    \includegraphics[width=0.95\textwidth]{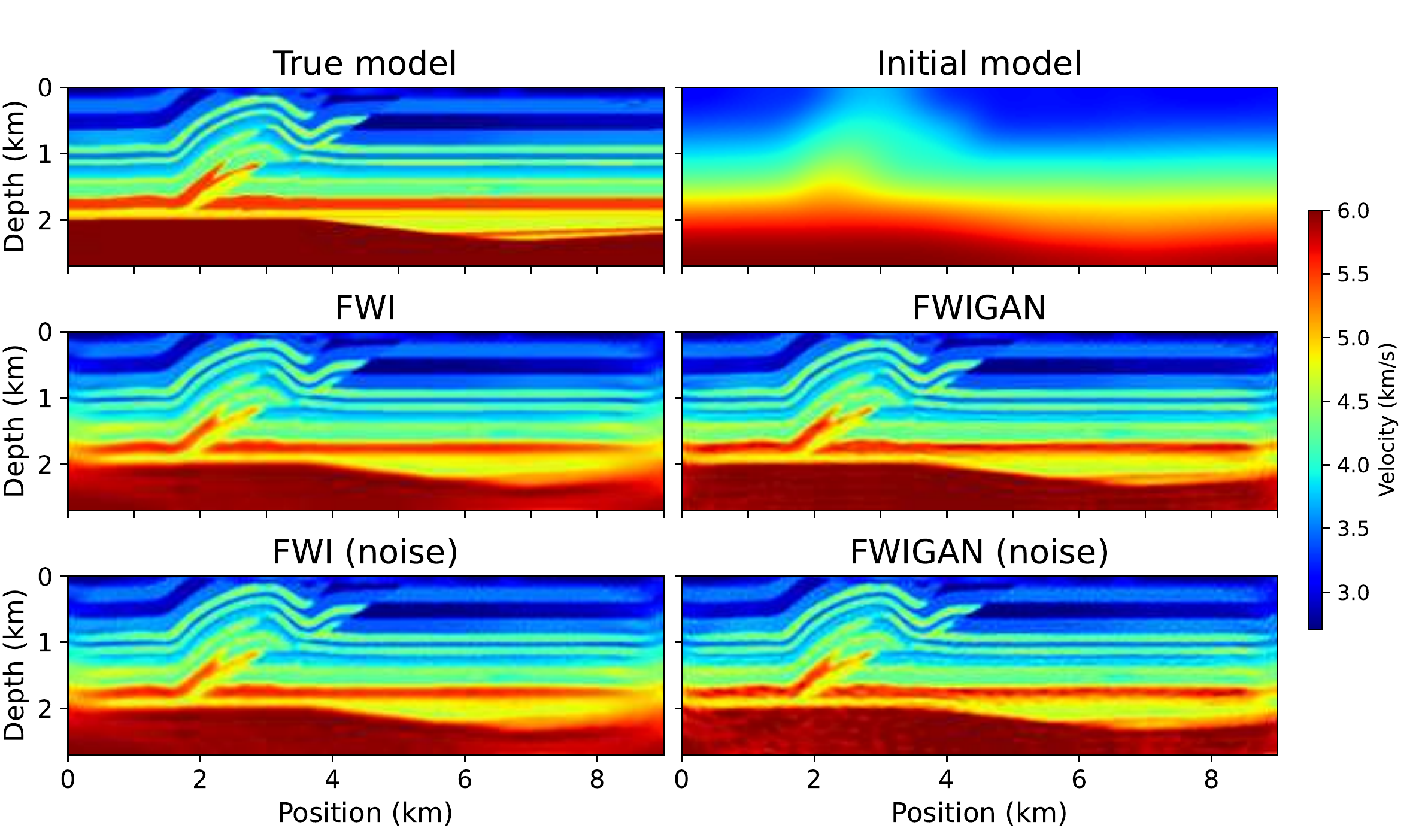}
    \caption{Comparable inversion results of Overthrust model. Frist row: true P-wave velocity model and initial guess. Second row: inverted results by FWI and the proposed method~(FWIGAN) with noise-free seismic data. Third row: inverted results by FWI and the proposed method~(FWIGAN) with noisy seismic data. The model size is $90 \times 300$ with a spacial interval of 0.03 km.
    }
    \label{fig:overv}
\end{figure*}

Additionally, we assessed how the proposed framework performs for Marmousi2 model. The performance of reconstructions with different approaches is shown in Figure~\ref{fig:mar2v}. Notably, the solution found by FWI exhibits inaccurate velocity values and smooth structures which do not fit the true model. On the contrary, FWIGAN generally obtains satisfying result that is consistent with ground truth. In particular, the structures below the sand at the center and the hull in deep layers.

Moreover, our method is more stable to obtain the expected features of geological structures even the seismic data is contaminated by noise. The quantitative metric shown in Table~\ref{tb:metric1} and Table~\ref{tb:metric2} and the estimated SNR verify this point as well.

\subsection{Results of Overthrust Model}

In Figure~\ref{fig:overv}, we give the inversion results by applying FWI and FWIGAN for Overthrust model. The SSIM and ERROR are also shown in Table~\ref{tb:metric1} and Table~\ref{tb:metric2} for both noise-free and noisy cases. Similarly, FWI leads to an acceptable result but our method offers a slight increase in the quality of the reconstruction for any configuration. These results suggest that the proposed scheme is robust to the complexity of geological structures and the missing information of deep layers. Meanwhile, it's able to attenuate the noise so that the denoised inversion results can be well produced.

\section{Discussion}\label{sec:discuss}

We introduced an robust and general  pipeline for FWI motivated by distribution matching. Based on our experiments, it appears that the results of FWIGAN are generally superior to those of FWI no matter the measurement is clean or noisy, which make the proposed framework of interest for practitioners.
To further demonstrate the flexibility and feasibility of FWIGAN, the discussion about estimation of source wavelet, sensitivity to different initial models,  and future works are  provided in this section.

\subsection{Estimation of source wavelet}
\begin{figure*}[tp]
    \centering
    \subfigure[]{\label{fig:mar1_noisefree}
    \includegraphics[width=0.48\textwidth]{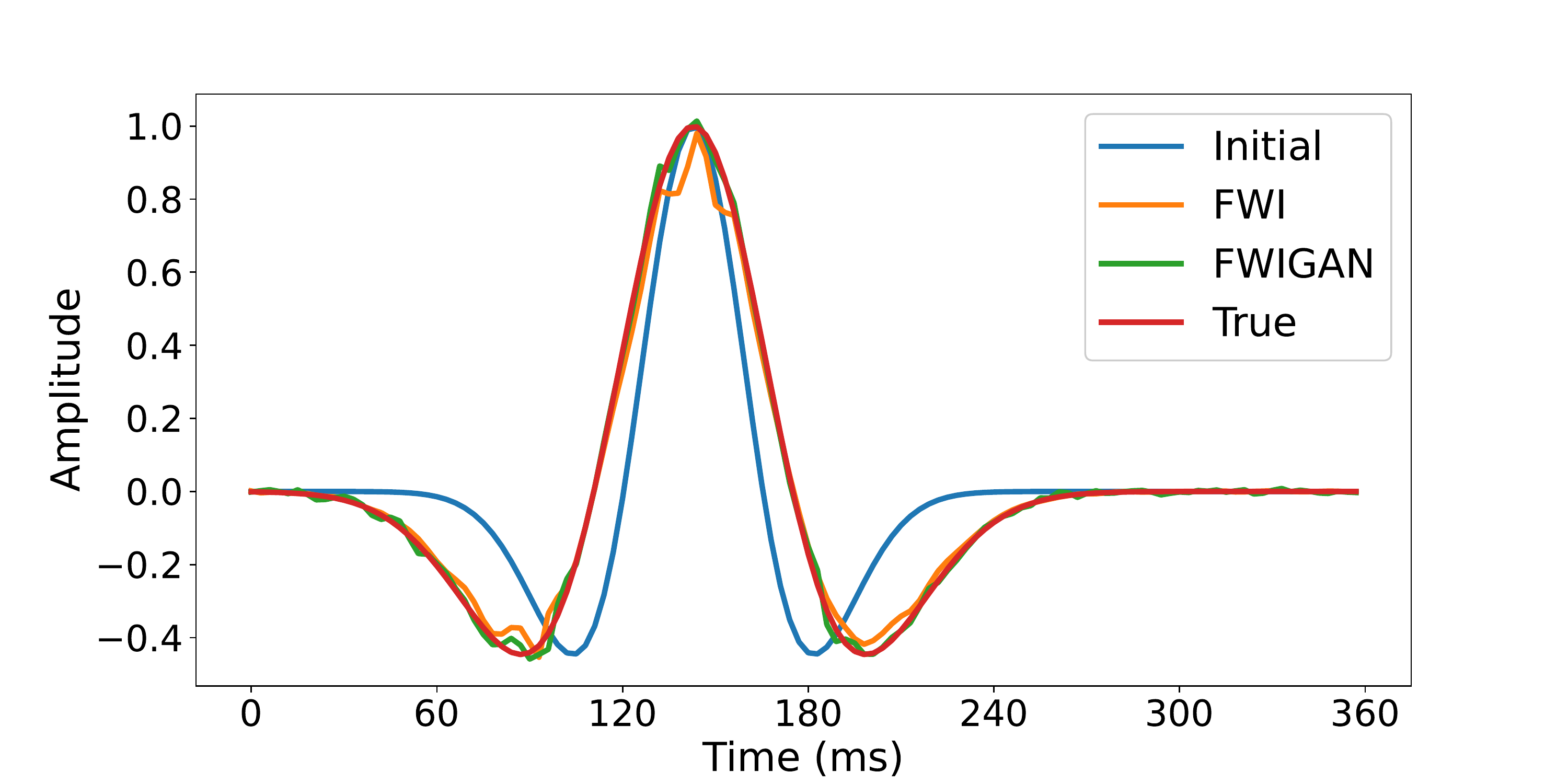}}
    \subfigure[]{\label{fig:mar1_noisy}
    \includegraphics[width=0.48\textwidth]{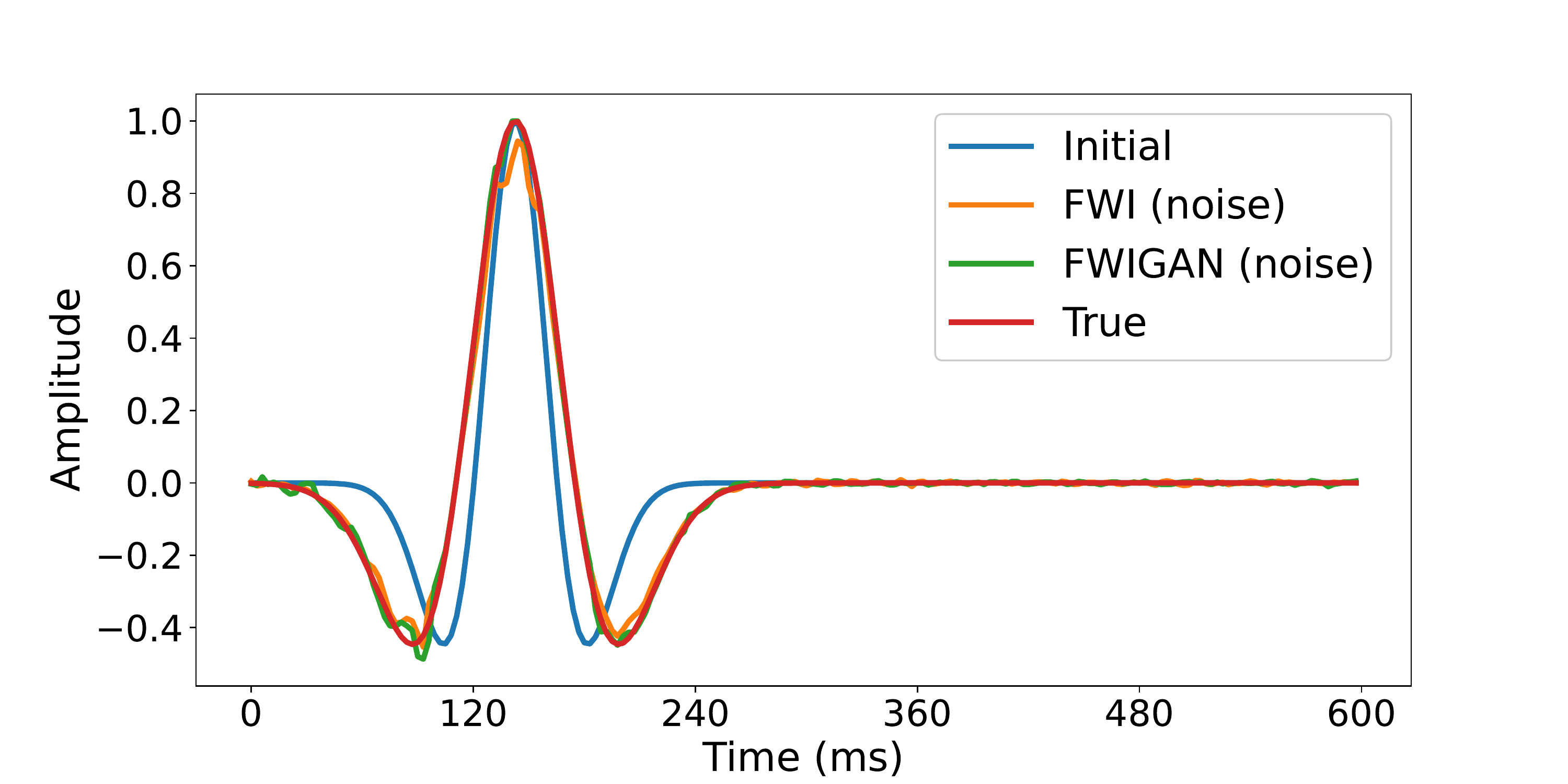}}\\
    \subfigure[]{\label{fig:mar2_noisefree}
    \includegraphics[width=0.48\textwidth]{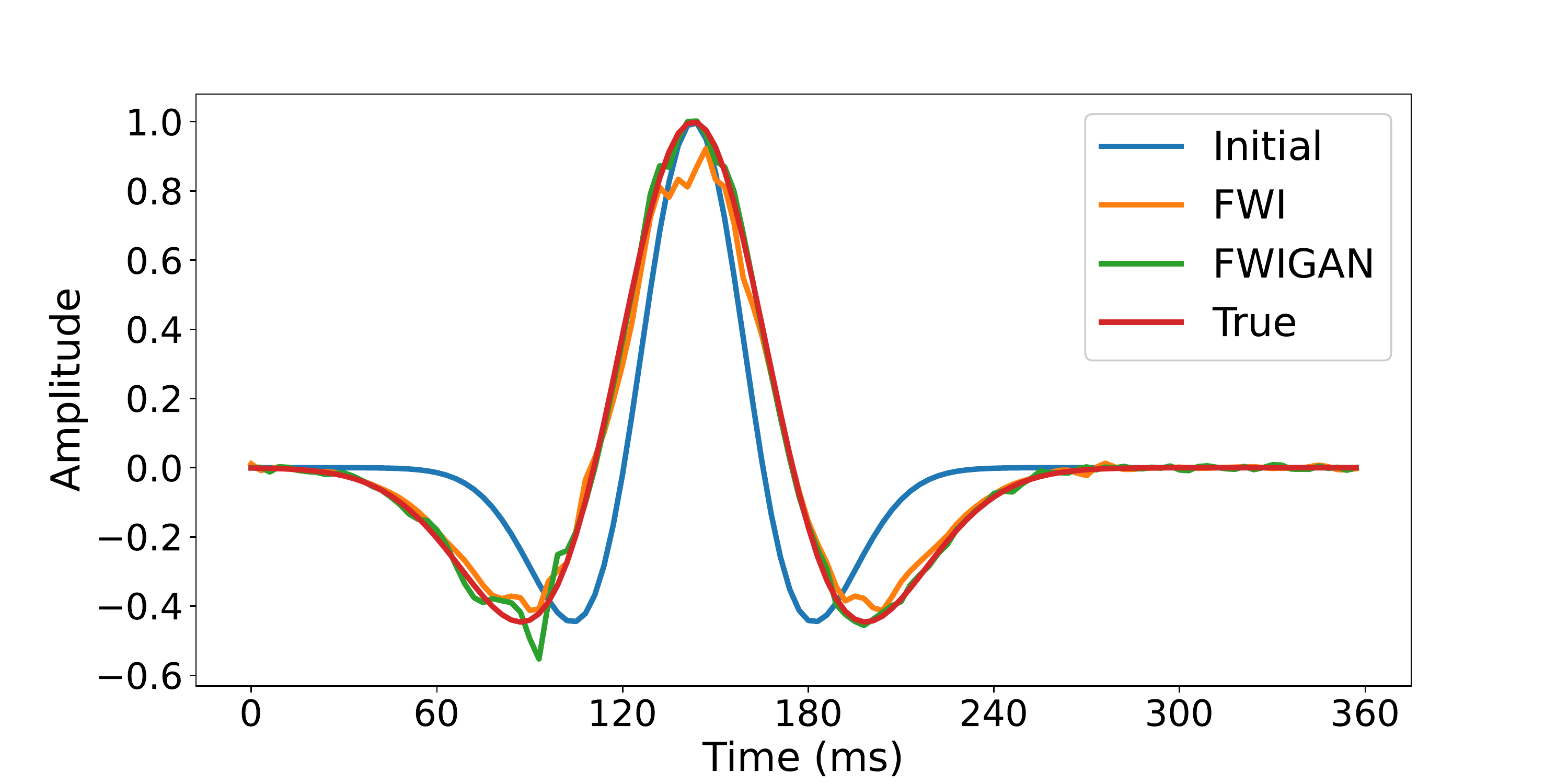}}
    \subfigure[]{\label{fig:mar2_noisy}
    \includegraphics[width=0.48\textwidth]{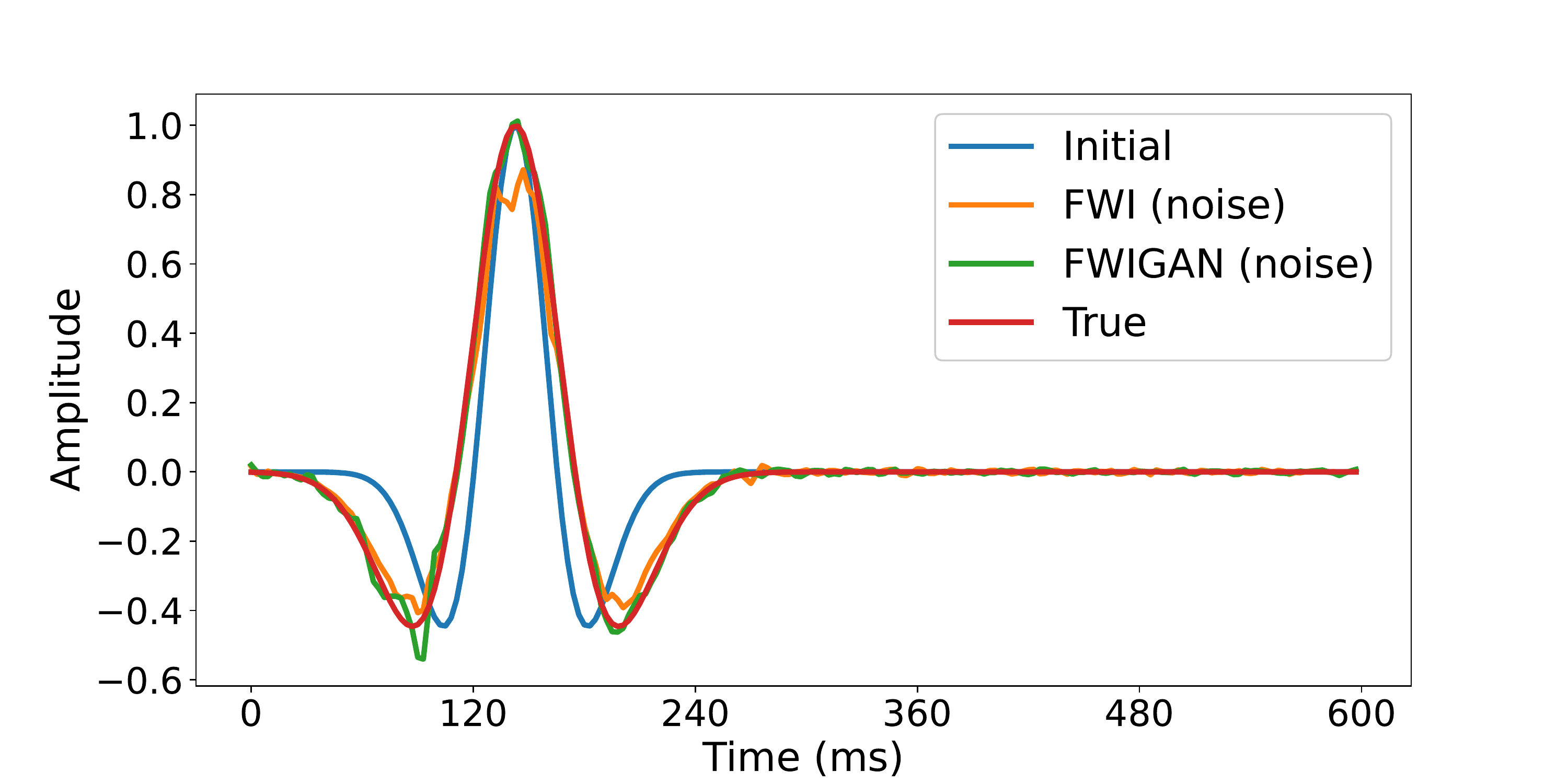}}\\
    \subfigure[]{\label{fig:over_noisefree}
    \includegraphics[width=0.48\textwidth]{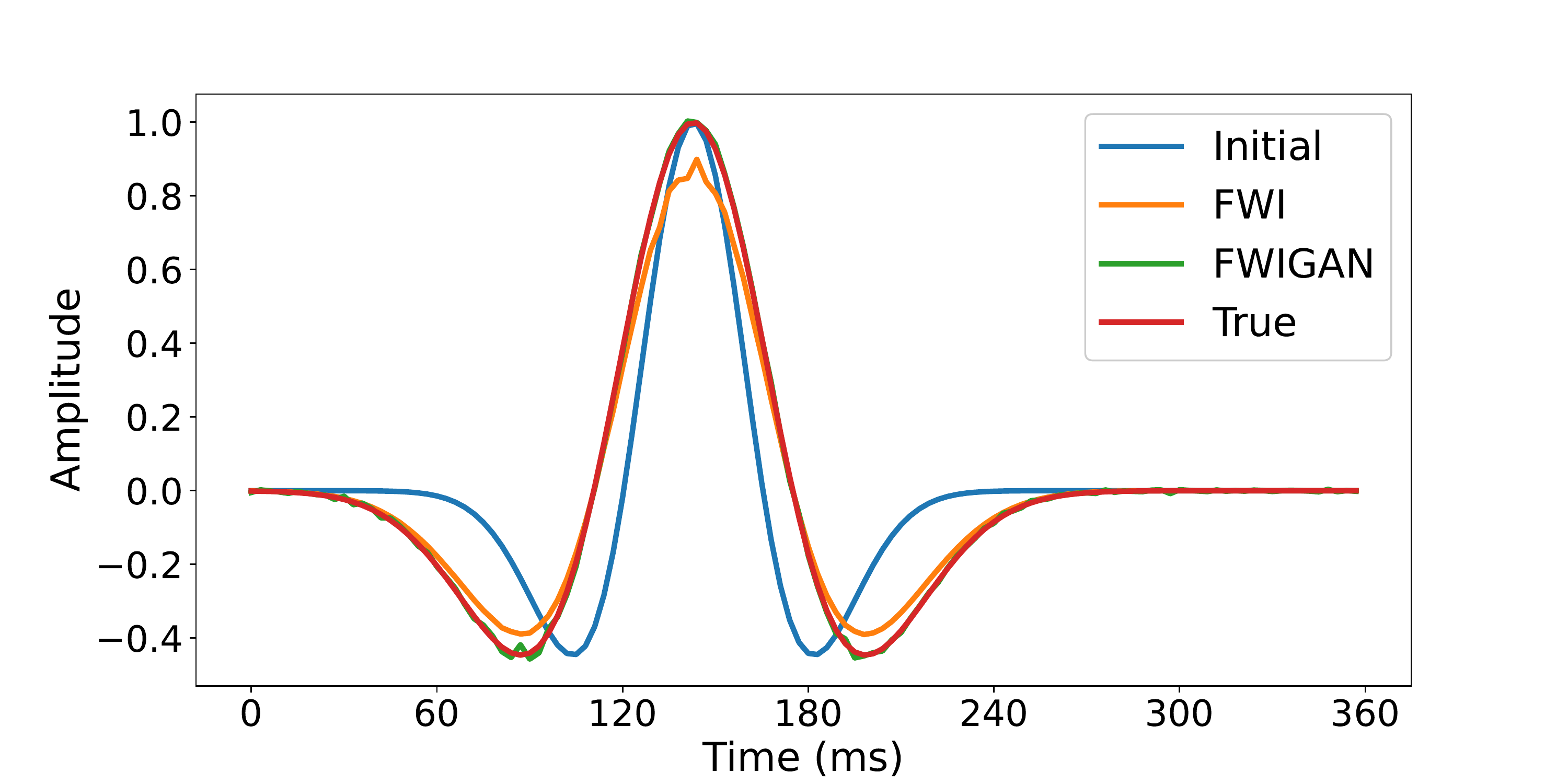}}
    \subfigure[]{\label{fig:over_noisy}
    \includegraphics[width=0.48\textwidth]{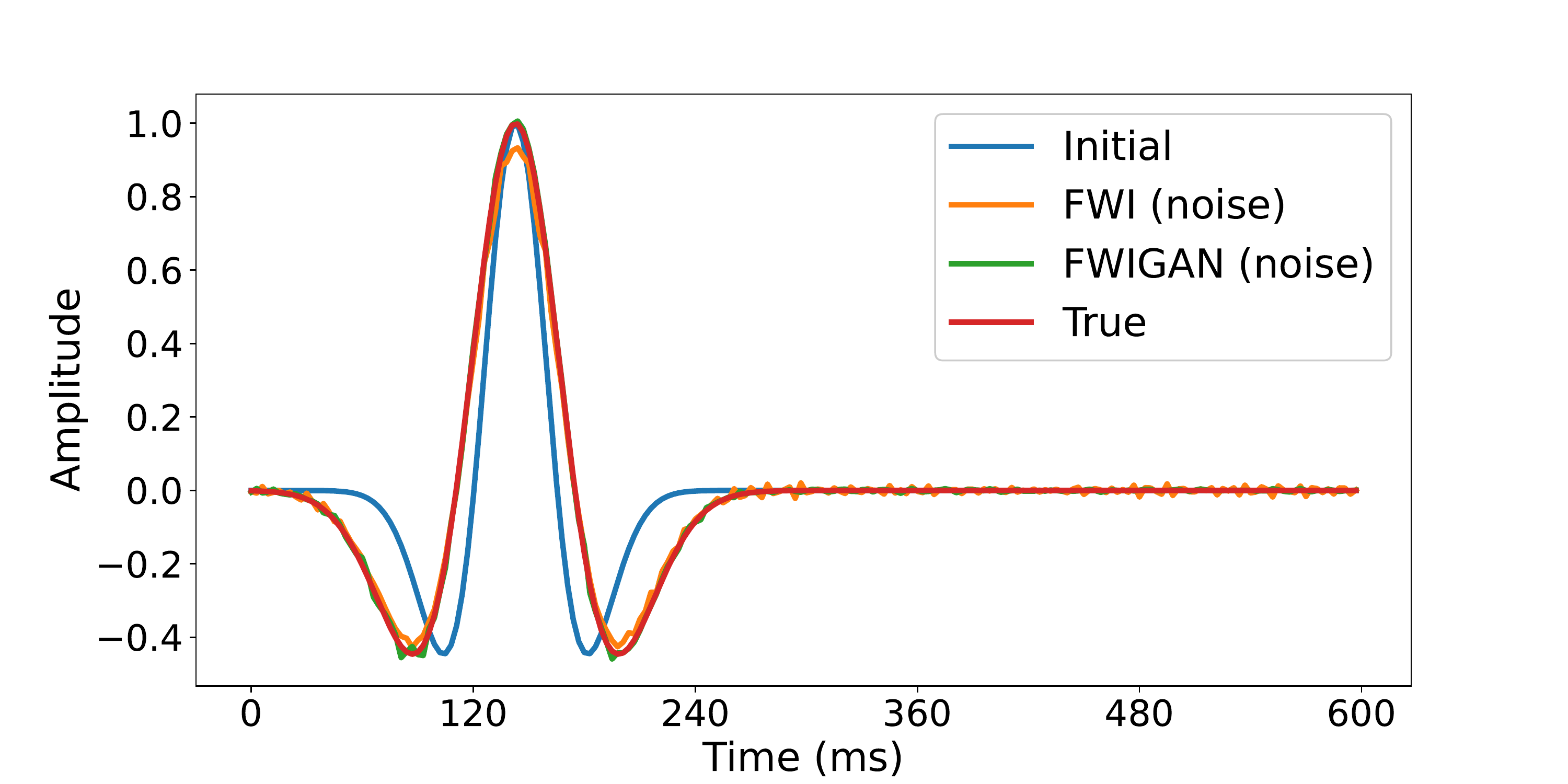}}
    \caption{Source wavelet estimation by FWI and FWIGAN with noise-free and noisy seismic data for Marmousi model (a)-(b), Marmousi2 model (c)-(d), and Overthrust model (e)-(f), respectively. The true and initial  source wavelet are same for these two configurations.}
    \label{fig:sourceinversion}
\end{figure*}

\begin{figure*}[tp]
    \centering
    \includegraphics[width=0.9\textwidth]{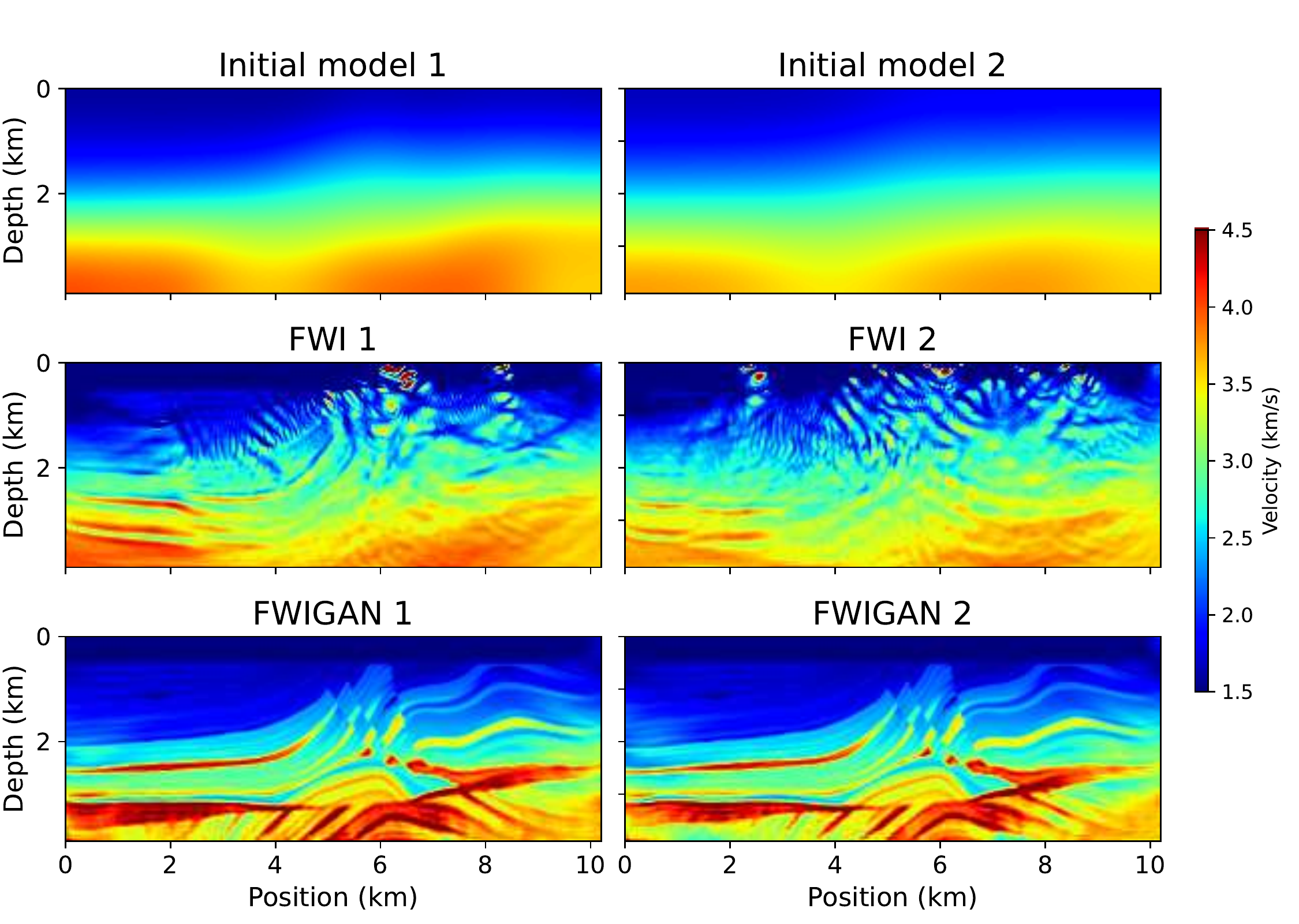}
    \caption{Inversion results of Marmousi2 model by FWI and FWIGAN with different smoothed initial model. Frist row: initial model obtained by Gaussian smooth function with true model whose standard deviation is 20 and 30, respectively. Second row: corresponding inversion results by FWI. Third row: corresponding inversion results by FWIGAN. The true model is shown in Figure~\ref{fig:mar2v}.}
    \label{fig:smooth}
\end{figure*}

As FWI applies a wave equation to delineate the subsurface velocity distribution according to the best fit between the simulated and observed data, the source wavelet estimation is one of the essential ingredients that may strongly
influence the seismic inversion results~\cite{hu2017demodulation}. Usually, the wavelet amplitude and phase spectra are estimated statistically from
seismic data. Last decades, a great number of works have been introduced to extract the source wavelet. In~\cite{frazer1998new}, the authors described a new objective function to invert sonic waveforms with an unknown source by using the receiver functions to interpret well-logged sonic data. Since the mapping between the source and the seismic wavefield is linear, the source function can be derived in FWI once the incident wavefield have been modeled. Then the medium and source are updated alternatively according to the cost~\cite{virieux2009overview}. To shy away the troublesome source inversion, developed frameworks have been designed both in frequency and time domains so the velocity inversion is independent from the source~\cite{lee2003source,choi2011source}. Another alternative of source wavelet estimation is to solve a nonlinear inverse problem by utilizing 1D wave equation and 2D prestack seismic data. This approach avoid any kind of seismic processing due to the source wavelet is the parameter to be reconstructed by optimization.
Aside from these methods, diverse new strategies are adopted to improve the performance.

In practical situations, the source signature of observed data can be obtained either by windowing a part of the early arrivals~\cite{yu2014application} or by predicting along with the velocity structure in the inversion algorithm.
However, it is hard to invert a good source wavelet without knowing the exact nearsurface velocity distribution. Therefore the assessed source signature can deviate fundamentally from the real source and may results in the final imaging is severely polluted with unacceptable artifacts.

Similar to reconstruct the velocity model by using the gradient-descent optimization strategy, the source wavelet can be forecast simultaneously in FWI workflow. It is worthy to note that inversion tasks of both velocity models and source wavelet is strenuous in FWI, for the sake of simplifying the inversion problem, we assume that the initial source guess is shifted in frequency from the true one and both true and predicted source wavelet are Ricker wavelet~\cite{ricker1951form}.
Consider the new inversion task, we can rewrite the cost function of classical FWI as
\begin{equation}\label{eq:FWIloss1}
    \min \mathcal{L}(\mathbf{v},f) =
    \min\frac{1}{2}\sum\limits_{s=1}^{n_s}\sum\limits_{g=1}^{n_r} \int_{0}^{T}|u_s(\mathbf{r}_g,t;\mathbf{v};f)-d_s(\mathbf{r}_g,t)|^2 dt.
\end{equation}
where $f\in \mathbb{R_{+}}$ is the peak frequency of the estimated source wavelet. Similarly, this variable will be adjusted like velocity model during the training process in the paradigm of FWIGAN.

For all experiments present in Section~\ref{sec:result}, we actually estimated the source wavelet at the same time. We used different learning rates for the model and source inversions as they have very different scales. In practice, the learning rate for source inversion was set as $10^{-3}$ and was decreased by the same strategy of model. The Adam optimizer was chosen for optimization. After the whole training, FWIGAN is able to return a reasonably inversion of the model and source. In Figure~\ref{fig:sourceinversion}, we show the source wavelet estimation performance of FWI and our method. By contrast, the proposed method leads to restoration that fit the true source better than that by FWI, especially the peak value. These results demonstrate that FWIGAN is capable of achieving multi-parameters inversion that general in piratical applications.

\subsection{Sensitivity to Initial Model}

While the aforementioned results are encouraging,
unfortunately, FWI at present can be attacked through local optimization algorithms so that building an accurate starting model for FWI remains one of the most topical issues.
To explore the sensitivity of FWIGAN to the accuracy of different
initial models, we extensively test the proposed framework with different smoothed models and linear model.

It is worthy to note that in our work, we attempt to introduce a novel paradigm that integrating deep learning with physics equation to achieve the task of seismic inversion even when the conditions are difficult, instead of improving the conventional optimization algorithms for FWI. Therefore the comparisons with developed approaches such as multi-scale frequency inversion for FWI with $\ell_2$ norm or other misfit function will be fixed in a future.

\subsubsection{Inversion Results with Smoothed Model}
\begin{figure*}[tp]
    \centering
    \includegraphics[width=1.\textwidth]{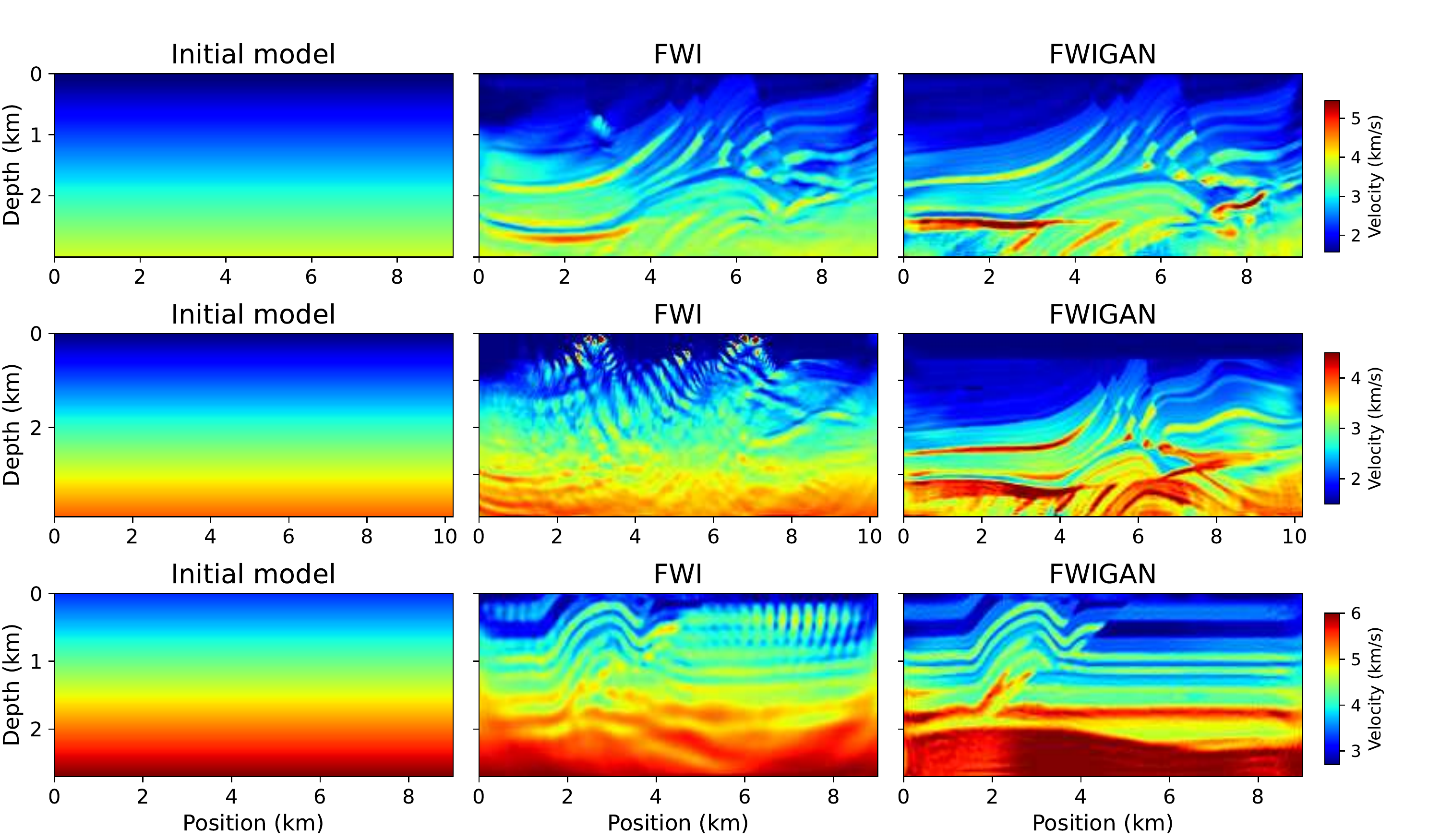}
    \caption{Inversion results by FWI and FWIGAN with linear initial model. Frist column: linear initial model of Marmousi, Marmousi2, and Overthrust models.  Second column: corresponding inversion results by FWI. Third column: corresponding inversion results by FWIGAN. The true models are shown in Figure~\ref{fig:mar1v}, Figure~\ref{fig:mar2v}, and Figure~\ref{fig:overv}, respectively.}
    \label{fig:linear}
\end{figure*}
We give in Figure~\ref{fig:smooth} the inversion performance of Marmousi2 model by applying FWI and FWIGAN. The starting model is built by Gaussian smoothed function with a higher standard deviation,  which is highly smoothed and far from the true model. Notably, FWI with $\ell_2$ norm dramatically underestimates the velocity distribution and fails to invert the accurate shape of the geological structures for these two cases. The results have spurious high-frequency artifacts which is a typical local-minima issue. On the contrary, our method obtains the high-quality reconstructions that match the true model better. It faithfully recovers most features of Marmousi2 model with lower ERROR and higher SSIM values (see Table~\ref{tb:metric3}).
These illustrative examples suggest that our proposed method can alleviate the trapping of local-minima despite the starting model is an undesired candidate.

\begin{table}[htbp]
\setlength{\tabcolsep}{8pt}
\centering
\caption{SSIM and relative error~(ERROR) of the inverted Marmousi2 model by using FWI and the proposed method~(FWIGAN) with different smoothed initial model.\label{tb:metric3}}
\arrayrulecolor{black}
\begin{tabular}{c | c | cc}
\toprule\toprule
Metric & Method  & Initial model 1  & Initial model 2 \\
\midrule
\multirow{2}{*}{{SSIM}}
&FWI    &0.1772 &0.1442 \\
&FWIGAN   &\textbf{0.7939} &\textbf{0.7567} \\
\midrule
\multirow{2}{*}{{ERROR}}
&FWI    &0.1870 &0.1929  \\
&FWIGAN   &\textbf{0.0789} &\textbf{0.1032} \\
\bottomrule\bottomrule
\end{tabular}
\end{table}

\subsubsection{Inversion Results with Linear Model}

Furthermore, we validate the effectiveness of the proposed framework with the linear initial model. The velocity value is increasing linearly in depth as $v(x,z) = v_0+\beta z$, where $v_0$ is the starting velocity value on the surface, $\beta$ denotes the average increments, and $z$ is the depth. As shown in Figure~\ref{fig:linear}, FWI produces blocky artifacts in all scenarios. It always fails to recover the correct structures, which points out the sensitivity of $\ell_2$ norm to the mismatch between the initial model and ground truth.
However, our method still recovers reconstructions with more accurate structures and velocity values. It performs well for most area except few parts at the bottom where are hard to precisely invert usually.

\begin{figure*}[htp]
    \centering
    \subfigure[]{\label{fig:mar1_linear}
    \includegraphics[width=0.25\textwidth]{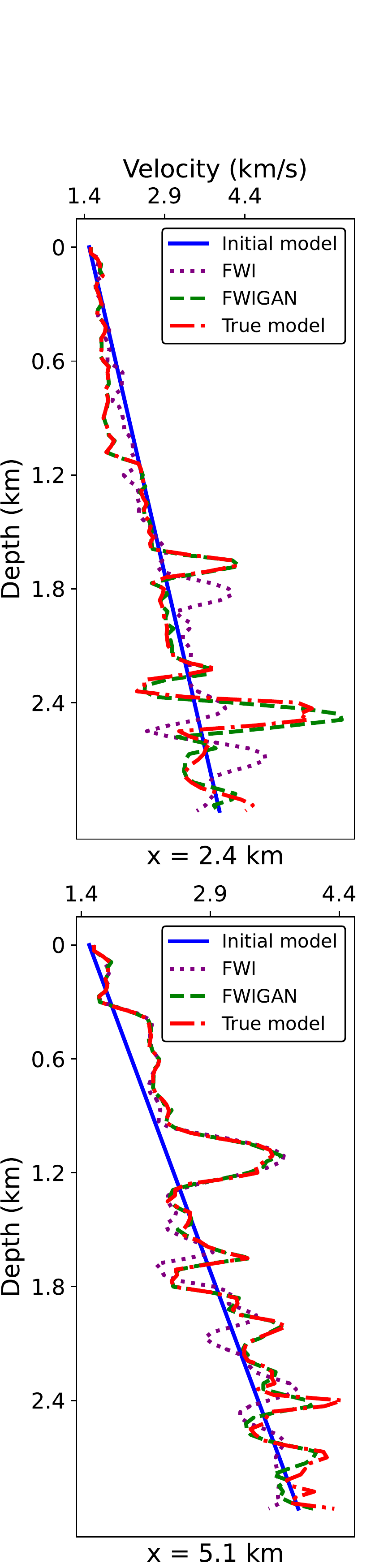}}
    \subfigure[]{\label{fig:mar2_linear}
    \includegraphics[width=0.25\textwidth]{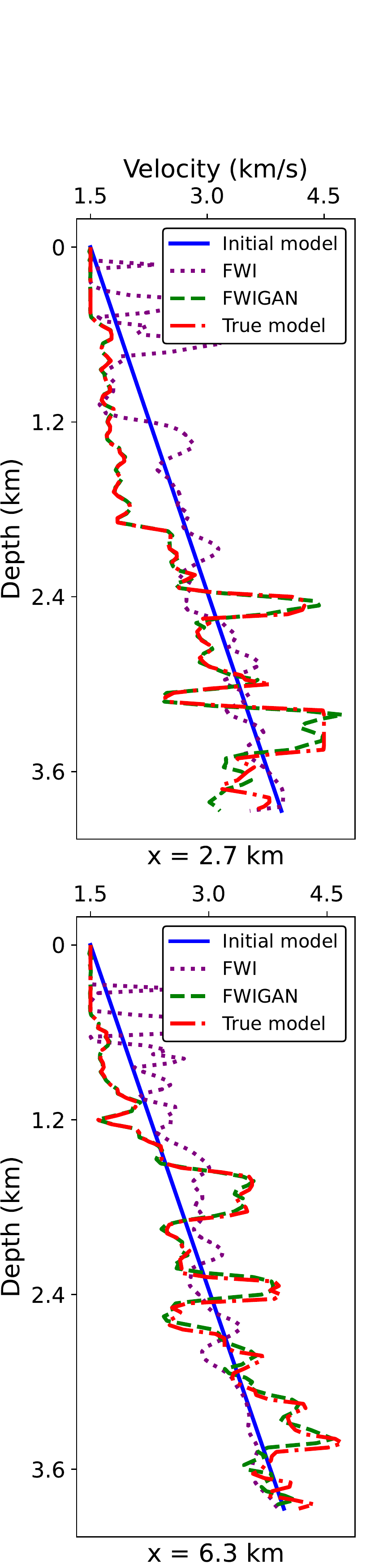}}
    \subfigure[]{\label{fig:over_linear}
    \includegraphics[width=0.25\textwidth]{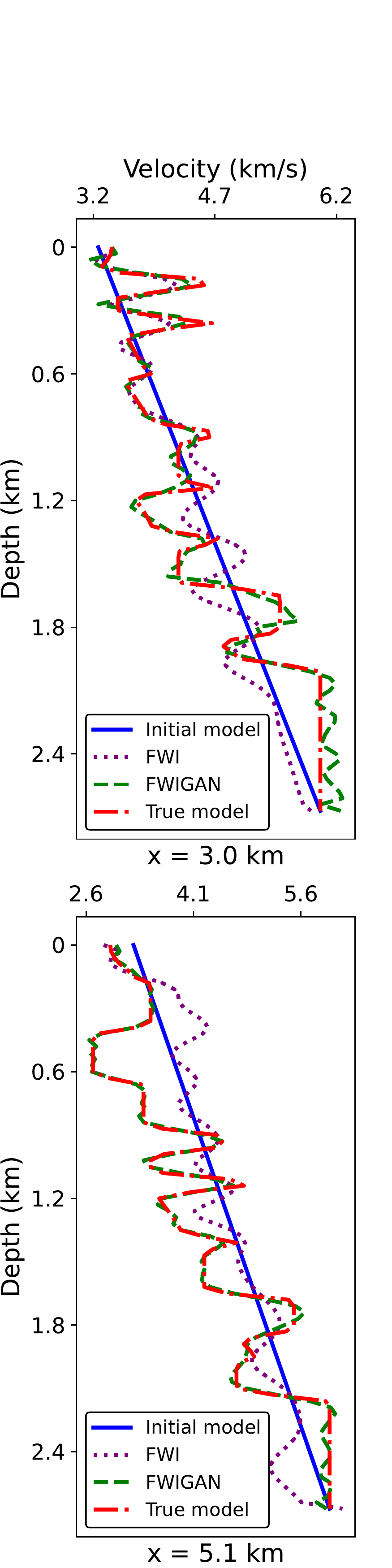}}
    \caption{The P-wave velocity profiles of true~(red), initial~(blue), FWI~(purple) and  FWIGAN~(green) inverted models at different horizontal locations. (a) Comparison of Marmousi model at position $x=2.4$ km and $x=5.1$ km. (b) Comparison of Marmousi2 model at position $x=2.7$ km and $x=6.3$ km. (c) Comparison of Overthrust model at position $x=3.0$ km and $x=5.1$ km.}
    \label{fig:verticalv}
\end{figure*}

For clear comparison,  the velocity profiles of true, initial, FWI, and FWIGAN inverted models at two horizontal locations for three models are visually illustrated in Figure~\ref{fig:verticalv}. By contrast, our proposed method leads to remarkable improvements such that the estimated velocities fit the true velocities better than those by FWI. The SSIM and ERROR measures shown in Table~\ref{tb:metric4} demonstrate our observation as well.

We emphasize that it may not perform well for all types of initial models, but these experiments show that the inversion by FWIGAN may still be satisfactory and reasonable when the starting model is far from the true model.

\begin{table}[htbp]
\setlength{\tabcolsep}{8pt}
\centering
\caption{SSIM and relative error~(ERROR) of the inverted velocity model by using FWI and the proposed method~(FWIGAN) with linear initial model.\label{tb:metric4}}
\arrayrulecolor{black}
\begin{tabular}{c | c | ccc}
\toprule\toprule
Metric & Method  & Marmousi  & Marmousi2 & Overthrust \\
\midrule
\multirow{2}{*}{{SSIM}}
&FWI    &0.3836 &0.1466 & 0.2690\\
&FWIGAN   &\textbf{0.7600} &\textbf{0.6527} & \textbf{0.7648}\\
\midrule
\multirow{2}{*}{{ERROR}}
&FWI    &0.1970 &0.1996 & 0.1137\\
&FWIGAN   &\textbf{0.1476} &\textbf{0.1281} & \textbf{0.0414} \\
\bottomrule\bottomrule
\end{tabular}
\end{table}

\subsection{Future Works}

Currently, the forward wavefield need to be saved in memory for use during back-propagation. It means the realistic 2D large-scale and 3D surveys will probably require more memory than is available. In addition, the design of the critic with complex architecture needs much memory as well. This computational limitation hinders the application of the proposed framework for practical models. The way to overcome this difficulty is one important direction of the future works.

Based on our experiments, we observe that the paradigms that rely on the DSL has the capacity of reconstructing multi-parameters owing to the AD and chain operations. Therefore it's easy to target at elastic full waveform inversion which is more accurate to simulate the wave propagation in isotropic data. However, the required storing memory for elastic wave equation is much more than that of acoustic wave equation, so the strategy to reduce the computational memory is a significant event.

Additionally, in terms of misfit function for FWI, adding regularization term is one usual way to improve the results and helps the optimization algorithm to overcome the local-minima problem. At present, we only adopt the Wasserstein-1 distance that is mostly used in DL field. For a future research, we will further explore different distance measure combined with regularization to tackle the difficulties of FWI.

\section{Conclusion}\label{sec:conclude}

We proposed a promising unsupervised learning paradigm FWIGAN that takes advantages of partial differential equations and deep learning techniques for two-dimensional seismic full waveform inversion. Motivated by the competitive training of Wasserstein generative adversarial learning, we integrate the acoustic wave equation with a neural network such that the physics generator is responsible for generating the physically constraint wavefield from current velocity estimation, and the critic discriminates the quality between observed and simulated data via distribution matching way. The goal of our framework is to recover the velocity model by making the distribution of simulated data as close as possible to that of observed data. It needs no ground truth nor pretraining of networks, and is flexible owing to the automatic differentiation so that requires minimal user interaction. We have validated our approach on well-known synthetic Marmousi, Marmousi2, and Overthrust models with diverse challenging conditions such as an undesired initial model and noisy measurements. Numerical experiments have demonstrated that the proposed method outperforms classical FWI algorithm with least-squares minimization in all configurations. Moreover, this model-data adjoint-learning pipeline is able to achieve multi-parameters inversion simultaneously in FWI workflow. The preliminary results suggest that our method is an appealing solution to alleviate the local-minima issues which make it of interest for practitioners.

\bibliographystyle{unsrt}
\bibliography{fwigan}  

\end{document}